\documentclass[11pt,a4paper]{article}
\usepackage[english]{babel}
\usepackage[a4paper,inner=2.5cm,outer=2.5cm,top=2.5cm,bottom=2.5cm]{geometry} 
\usepackage[square,numbers]{natbib} 
\usepackage{amsmath} 
\usepackage{enumitem}
\usepackage{parskip}
\usepackage{graphicx}
\usepackage{bbm}
\usepackage{subcaption}
\usepackage{caption}
\usepackage{wrapfig}
\usepackage{adjustbox}
\usepackage{tabularx}
\usepackage{array}
\usepackage{hyperref}
\usepackage[toc,page]{appendix}
\usepackage{algorithm}
\usepackage{algpseudocode}

\algnewcommand\algorithmicforeach{\textbf{for each}}
\algdef{S}[FOR]{ForEach}[1]{\algorithmicforeach\ #1\ \algorithmicdo}
\newcolumntype{L}[1]{>{\raggedright\let\newline\\\arraybackslash\hspace{0pt}}m{#1}}

\title{A Diagnostic to Find and Help Combat Stochastic Positivity Issues -- with a Focus on Continuous Treatments}
\date{}
\author{
  Katharina Ring$^{1}$\thanks{Corresponding author. Contact: katharina.ring@stat.uni-muenchen.de},
  Michael Schomaker$^{1,2}$ 
}
\begin{document}

\maketitle

\vspace{-2em}
\begin{center}
      $^1$Department of Statistics, Ludwig-Maximilians University Munich, Geschwister-Scholl-Platz 1,
80539 Munich, Germany \\
  $^2$Centre for Infectious Disease Epidemiology and Research, University of Cape Town, 7925 Cape Town, South Africa
\end{center}

\section*{Abstract}

The positivity assumption is central in the identification of a causal effect. Especially its stochastic variant is an issue many applied researchers face. Yet positivity is rarely discussed, especially in conjunction with continuous treatments or Modified Treatment Policies. One common recommendation for dealing with a violation is to change the estimand. However, an applied researcher is faced with two problems: 
First, how can she tell whether there is a stochastic positivity violation given her estimand of interest, preferably without having to estimate a model first?
%This is especially challenging for continuous treatments, where simply checking cross tables is not possible.
Second, if she finds a problem with stochastic positivity, \textit{how} should she change her estimand in order to arrive at an estimand which does not face the same issues? %Ideally, this new estimand should also be as close as possible to her original research question.
We suggest a novel diagnostic which allows the
researcher to answer both questions by providing insights into how well  an estimation for a certain estimand can be made for each observation using the data at hand. %This diagnostic can be applied to several estimands in order to find the best estimable one. 
%A researcher may choose the particularly large class of Modified Treatment Policies (MTP), which relax the assumption of positivity to some extent, in order to produce a rich set of viable estimands.
We provide a simulation study on the general behaviour of different Modified Treatment Policies (MTPs) at different levels of stochastic positivity violations and show how the diagnostic helps understand where  bias is to be expected. We illustrate the application of our proposed diagnostic in a pharmacoepidemiological study based on data from CHAPAS-3, a trial comparing different treatment regimens for children living with HIV.

\section{Introduction}

 Even though the assumption of positivity is essential to most standard causal effect estimation procedures, it is rarely discussed. In addition, most of the few works that consider consequences and solutions to positivity violations in detail are heavily focused on the binary treatment case, for example because ``[...] it is the most commonly explored setting in the existing literature and has a well-developed set of methods available for comparison.'' \cite[page 1474]{zhu2021core}. However, continuous treatments are very often of interest in practice \cite{kennedy2017non}. Theoretical positivity violations can often be identified using logic, as one of the typical examples, an intervention where the treatment hysterectomy is assigned to a man without a womb \cite{westreich2010invited}, illustrates.
Stochastic positivity violations on the other hand can be detected via cross tabulation in sufficiently simple binary settings. But diagnosing them in a continuous setting is not trivial as there are always more values the variables can take than there are data points. 
 This shows the need for a diagnostic that helps an applied researcher detect stochastic positivity violations specifically in continuous and high-dimensional data.
Once a researcher has reached the conclusion that there are stochastic positivity issues with an estimand, a common recommendation is to switch the estimand \cite{zhu2021core}. Ideally, the researcher will find a compromise making sure the estimand is as close as possible to the original research question and avoiding bias due to positivity issues \cite{petersen2012diagnosing}. But this reveals another problem: How can a good alternative estimand, which solves this trade-off well, be found?

Regarding the identification of stochastic positivity issues, some tools exist which can be used in continuous treatment cases. Petersen et al.\ \cite{petersen2012diagnosing} suggested a parametric bootstrap which, given an estimator, produces an optimistic estimate of the bias. This tool may not always be very accurate and captures not only positivity issues but also other forms of sparsity, rendering its main purpose serving as a warning sign. The main drawback for this diagnostic is that it is estimator-specific. It can only be used once an estimator has been chosen and does not answer the question if the issue is with the estimator or the data. In addition, once stochastic positivity issues are identified with this tool, there is no guidance on what can be done to combat this issue. Another approach suggested by Zhu et al.\ \cite{zhu2021core} is to generalize propensity score based approaches, which are often used in order to deal with stochastic positivity issues in binary treatments \cite{10.1093/biomet/asn055}, to continuous treatments. The generalized propensity score (GPS) is an estimate of the conditional density of the treatment given the variables of the adjustment set. However, this requires modeling a density first and leaves the researcher with a potentially high-dimensional density estimate, which cannot easily be used to derive information about positivity. In addition, the generalized propensity score assumes that observations with every feature combination require positivity in all potential treatments, but this is only true for a static treatment scheme. As discussed before, changing the estimand, typically to a more elaborate scheme, is one of the solutions to positivity problems, therefore a useful diagnostic needs to work on a large variety of estimands. Finally, the possibility of extrapolation from high density regions is not addressed at all if stochastic positivity is diagnosed via the GPS.
Recently, Danelian et al.\ \cite{DanelianFoucherLégerLeBorgneChatton+2023} proposed a regression tree based method to identify sparse areas called positivity regression tree (PoRT) algorithm. Chatton et al.\ \cite{chatton2024regressiontreesnonparametricdiagnostics} extended this to longitudinal settings with the sequential PoRT (sPoRT) algorithm. So far the authors have only considered binary exposure, but the algorithm could be adjusted to other types of treatment. However, as the algorithms depend on hyperparameters to define what constitutes a stochastic positivity issue, an extension to continuous treatments would mean this non-trivial question has to be answered first, likely falling back to ideas like the GPS and some threshold value.

%[todo] tree algorithm Michael told me about?

Once a researcher has identified a stochastic positivity problem and wishes to change the estimand, they can use one of several estimands with relaxed positivity assumptions. For example, Kennedy \cite{kennedy2019nonparametric} introduced a class of estimands called incremental propensity score interventions which do not rely on any positivity assumption as the propensity score itself is used to define the estimand. While this is potentially an interesting estimand to a researcher facing positivity issues, it solves the trade-off between bias due to stochastic positivity issues and closeness to the estimand of interest rather one-sidedly. In addition, the estimand is not necessarily easy to interpret and may not lead to a meaningful conclusion within the research field  in question. Another large class of estimands with relaxed positivity assumptions are modified treatment policies (MTP) \cite{diaz2023nonparametric}. While MTP estimands leave the researcher a lot of options to define an estimand that can likely be estimated without suffering from a high bias due to stochastic positivity issues, it remains unclear which MTPs will result in a satisfactorily low bias.

In order to help applied researchers diagnose stochastic positivity issues and find effective ways to respond to them, we introduce a diagnostic in this work. The diagnostic shows, given an estimand, how much support there is for all $n$ prediction problems of the causal estimand and helps a researcher compare different estimands based on which are likely to avoid bias due to stochastic positivity issues given a data set. The diagnostic can also act as a guide to a researcher aiming to tweak an estimand in order to avoid bias due to stochastic positivity issues.

In the next section, the notation and setup of the work is introduced as well as core concepts. In Section \ref{diagnostic}, the diagnostic is first motivated and then two versions are introduced alongside a guide on how to choose hyperparameters. Section \ref{chap_simulation} contains a simulation study where the diagnostic is applied to different MTPs estimated on data with varying degrees of stochastic positivity violations, highlighting the issues a researcher faces when trying to choose an alternative estimand from this class. The diagnostic is applied to medical data in Section \ref{actualdata}, where it is demonstrated how the diagnostic can be used to tweak problematic estimands. The paper concludes with a discussion of advantages and limitations in Section \ref{discussion} and a conclusion in Section \ref{conclusion}.

\section{Theoretical Background}

\subsection{Notation and Setup}

We consider an observed sample of $n$ independent and identically distributed (i.i.d.) observations of the form $O = (L,A,Y) \sim P_0$, where $L$ encompasses all adjustment variables needed to recover the causal quantity of interest, $A$ is the discrete or continuous treatment which we wish to intervene on, and $Y$ is the outcome variable. While the focus of this work lies on a standard outcome and univariate treatment, the diagnostic, which we introduce later on, can easily be applied to, e.\ g., survival outcomes, or multivariate $A$ as well. 
Following the potential outcomes framework \cite{rubin1974estimating}, we denote the counterfactual outcome $Y$, which we would have observed had the treatment value been $a$, as $Y^a$. Instead of a fixed value $a$, corresponding to a static intervention, the treatment could also be determined by a function. For an overview of different types of intervention schemes see Section \ref{lmtp}. In order to differentiate more easily between the observed sample and the intervened on sample, i.\ e., the sample with a changed value $a^{int}$ for $A$ according to the intervention scheme, we denote the observed values for a subject $i$ as $o_i^{obs}=(l_i, a_i^{obs})$ and the corresponding counterfactual values as $o_i^{int}=(l_i, a_i^{int})$. The outcome $y_i$ is omitted here as it is not used in the diagnostic we present.

Given the three assumptions of consistency, (theoretical) positivity and unconfoundedness are fulfilled,
there are multiple ways to recover a causal quantity of interest $\psi(P_0)$ \cite{petersen2012diagnosing}. The focus of this work lies on three popular estimators, which we introduce here using the example of estimating the expected outcome of $Y$ under the intervention strategy. Firstly, g-computation can be employed, which requires the estimate $\bar{Q}_n$, a model of $\mathbbm{E}(Y|A,L)$. Secondly, inverse probability of treatment weighting (IPTW) can be utilized, which requires an estimate $g_n$, a model of the probability of treatment $P_0(A=a|L)$ in the binary case or, respectively, $f_{0,A|L}(A|L)$ in the continuous case (the subscript zero will be dropped from the density subsequently for better readability). Lastly, doubly robust methods require the estimation of both. While other methods exist, they are less general, typically either specialized to specific data structures (e.\ g., instrumental variable approaches \cite{zeng2025nonparametricestimationlocaltreatment}), rather basic (e.\ g., stratification), or take a completely different point of view on estimation, like some Bayesian approaches \cite{nethery2019estimating}, therefore we do not consider them in this work.

For simplicity, the main focus of this paper is on single time point analyses. However, much of this work can easily be extended to multiple time point causal quantities. We give some intuition on that in Section \ref{outlook:longi}.

\subsection{Positivity}
\subsubsection{Definition}

% use int obs notation here as well? But I guess it is not necessary yet
The positivity assumption for a continuous treatment is typically stated as
%For a continuous treatment and a static intervention scheme the positivity assumption is given by 
$$f_L(l)\neq 0 \Rightarrow f_{A|L}(a|l)>0 \,\, \forall a,l,$$
referring to a range of static interventions typically depicted in a causal dose-response curve \cite{hernan2023causal}. In Section \ref{lmtp}, this definition is extended to other intervention schemes.
%which extends to
%$f_{\bar{A}_{k-1},\bar{L}_k} (\bar{a}_{k-1}, \bar{l}_k) \neq 0 \Rightarrow f_{A_k|\bar{A}_{k-1},\bar{L}_k} (a_k|\bar{a}_{k-1}, \bar{l}_k) >0 \,\, \forall (\bar{a}_k,\bar{l}_k) $
%in the longitudinal case \cite{hernan2023causal}.
% positivity, according to the literature, can be split into two different types which can then again be split in two different types
This positivity assumption postulates that, given possible covariate combinations, it is possible for an observation to receive all treatments of interest. 
The positivity assumption has been split into two different types: deterministic and stochastic positivity \cite{zhu2021core, petersen2012diagnosing}. Deterministic, or theoretical, violations encompass especially illogical estimands, for example, a pregnancy in a person without a female reproductive system, and are therefore not an estimation issue per se. Stochastic, or practical, violations on the other hand cover issues, where for a certain subgroup of observations no or very little data are observed for some values of treatment, either due to low probabilities in the data generating process or simply chance. Stochastic positivity violations result in sparsity in the finite sample, which poses a challenge in estimation as they go hand in hand with a high bias \cite{petersen2012diagnosing}. Note that we use the term sparsity in this work not as a descriptor for a problem stemming from high-dimensionality of the data alone, but as an issue related to local sample size, which is influenced by the sample size, the dimensionality of the data, and the distribution of the data.

%\begin{enumerate}
%\item deterministic positivity violations
%\begin{enumerate}
%\item strictly deterministic: illogical treatments like negative doses of medication
%\item structural: highly unlikely, e.\,g., due to contraindication or illegality
%\end{enumerate}
%\item stochastic positivity violations
%\begin{enumerate}
%\item full violation: by chance no observation with $L=l$ received treatment $a$
%\item near violation: by chance few observations with $L=l$ receive treatment $a$
%\end{enumerate}
%\end{enumerate}
%\cite{westreich2010invited, zivich2022positivity}
% actually maybe even structural violations are not really positivity violations
%A deterministic violation always implies a stochastic violation and therefore methods for deterministic violations can always be applied in the presence of stochastic violations but not vice versa \cite{zivich2022positivity}.
% what are the ways to deal with deterministic?

%Stochastic violations can be seen as data sparsity problems \cite{petersen2012diagnosing} and are therefore a problem of estimability \cite{zivich2022positivity}. 
\subsubsection{Responding to Stochastic Positivity Violations}
Current recommendations for dealing with sparsity due to stochastic positivity violations are extrapolation (in fact, if perfect extrapolation can be assumed, the assumption of positivity is not needed anymore \cite{zhang2025nonparametricinferencedoseresponsecurves}), if possible, and/or changing the estimand \cite{zhu2021core, petersen2012diagnosing}. The former requires using an estimator which allows for extrapolation \cite{petersen2012diagnosing}. The latter can be approached either by adjusting the target population (trimming/weighting the sample) \cite{zhu2021core, petersen2012diagnosing}, or by adjusting the intervention scheme of interest, both of which represent a trade-off between proximity to the original research question of interest and identifiability of the effect \cite{petersen2012diagnosing}.
An alternative to extrapolation and changing the estimand is restricting $L$, resulting in a trade-off between bias due to stochastic positivity issues and bias due to confounding \cite{petersen2012diagnosing}. Within this work we will address all of these proposed solutions to stochastic positivity violations. 
The relation of stochastic positivity to inter- and extrapolation, which is a necessity in the continuous case, will be considered in detail in Section \ref{diagnostic}, where we introduce our diagnostic. The diagnostic provides guidance on adjusting the target population, choosing a potential variable from the adjustment set to exclude from the analysis, and deciding between several potential estimands. However, in order to create alternative estimands, one has to be able to choose from a large variety of schemes in order to find the one providing the best trade-off between meeting your needs as a researcher and being estimable with the data at hand. Therefore, in the following we provide an overview of several types of intervention schemes, including Modified Treatment Policies (MTPs), which are a large class of intervention schemes from which several estimands can be created.

%------------------------------------------
% THE LMTP
%------------------------------------------

\subsection{Advanced Treatment Schemes and Modified Treatment Policies}\label{lmtp}

The simplest and most often used intervention scheme is the static intervention $A=a$, where $a$ is the same fixed value for all subjects in the study. Sometimes many values for $a$ are of interest and the goal is to estimate a Causal Dose Response Curve (CDRC), where for each value of $a$ the counterfactual outcome $Y^a$ is presented in a graph \cite{kennedy2017non}. For this type of intervention the strict positivity assumption as stated above needs to be met.
The class of static interventions can be extended to dynamic interventions, where the value of the treatment depends on variables of the adjustment set $l$ leading to $A=\mathbbm{d}(l)$, for some function $\mathbbm{d}$ \cite{diaz2023nonparametric}. The positivity assumption for a dynamic scheme is less strict compared to a fixed scheme. For example, if a certain treatment value $a$ only occurs for a sub-group of the sample with $L=l$ in the estimand, $f(a|L)$ only needs to be positive for $L=l$. Modified Treatment Policies (MTP) extend dynamic schemes by letting the function $\mathbbm{d}$ depend on the natural value of $A$, leading to $A=\mathbbm{d}(l, a^{obs})$ \cite{diaz2023nonparametric}. This relaxes the positivity assumption further. 
%For Longitudinal MTPs, the formula is given by $A_t=f(A_t^{nat}, H_t^{int})$, where $H_t^{int}$ is the history of the subject under intervention at time point $t$. 
Doubly robust estimation of MTP causal estimands as described by D{\'\i}az et al.\ \cite{diaz2023nonparametric} can be conducted using the R-Package lmtp \cite{williams2023lmtp}.
%\citet{diaz2023nonparametric} also introduce a stochastic version of MTP to be utilized in cases where a fixed assignment of treatment may be unrealistic. %The stochastic variant also has the advantage of relaxing the conditional exchangeability assumption compared to a stronger assumption needed for MTP. (I think this is just in the longitudinal case)

The positivity assumption can be stated more generally, encompassing all types of intervention schemes named here. For a single time point this assumption is given by 
\begin{align*}
f_{A,L}(a^{\text{obs}},l) \neq 0 \Rightarrow f_{A,L}(a^{int}|l) > 0 \,\,\forall (a^{int}=\mathbbm{d}(a^{\text{obs}}, l),l) \in E,
\end{align*}
where $E$ are combinations of $a^{int}$ and $l$ which are seen in the estimand (see Appendix \ref{extendedpositivity}). For example in a dynamic scheme, where women are given a dose of 5\,g of medication and men receive 3\,g and no other variables exist, positivity is only needed for $f(a=5|\text{woman})$ and $f(a=3|\text{man})$. Therefore MTPs relax the positivity assumption by giving the researcher more freedom to define a treatment scheme which avoids sparse data regions and represents an important tool in the toolkit of every researcher trying to combat stochastic positivity issues.

%------------------------------------------
% DIAGNOSTIC
%------------------------------------------

\section{A Kernel-based Sparsity Diagnostic}\label{diagnostic}

In this section, we introduce a feasibility diagnostic, which provides guidance to applied researchers about whether a causal estimand can, in principle, be estimated given a specific data set. The diagnostic is designed to be \textit{estimand-specific}, because the positivity assumption depends on the estimand, and \textit{data set-specific}, because stochastic positivity violations are a property of the data, not necessarily the data generating process. The diagnostic is also designed to be \textit{estimator-independent}, since as long as there is one estimator that can recover the causal quantity of interest, the research question can be answered. In order to construct this diagnostic, the sources of bias related to positivity issues in estimation are examined in the following.

\subsection{Sources of Bias}\label{sourcesofbias}
Petersen et al.\ \cite{petersen2012diagnosing} identify four sources of bias in the estimation procedure for a causal estimand:
\begin{enumerate}
    \item Estimators for $g_n$ and/or $\bar{Q}_n$ are inconsistent
    \item $g_0$ violates the positivity assumption
    \item Finite sample bias for consistent estimators of $g_n$ and/or $\bar{Q}_n$
    \item Estimated values of $g_n$ are close to zero (if applicable)
\end{enumerate}

% general bias: gn Qn inconsistent, g0 positivity violated, gn Qn finite sample bias, gn may be close to zero
% positivity bias: g0 positivity violated, truncation as response to gn may be close to zero, finite sample bias because gn close to zero
% sparsity bias: 
The first source of bias is often addressed by using doubly robust estimation with a large variety of highly adaptive models in the super learner \cite{van2011targeted}. This is the only source of bias named by Petersen et al.\ which is not linked to positivity and is therefore not addressed further. The second source of bias represents the deterministic positivity violation and means there are areas in the feature space where it is impossible to observe data. This can be mitigated using inter- and extrapolation from neighboring areas where data was observed, albeit by introducing another assumption, for example in the form of a parametric model. 
Inter- and extrapolation are also a necessity when dealing with the third form of bias, finite sample bias for $g_n$ and $\bar{Q}_n$, at least in a continuous setting, as a finite sample means there is no data almost everywhere.
Finite sample bias is an issue in any finite data set, but the problem is exacerbated in sparse regions of the data. Therefore, practical positivity violations lead to an increase in finite sample bias. In a continuous setting, both 2 and 3 force us to rely on inter- and extrapolation, the only relevant difference in estimation between the two is that 3 could theoretically be mitigated by collecting more data. For a given data set, however, they can be treated as the same problem: sparsity \cite{petersen2012diagnosing}. The last source of bias stems from low estimates of $g_n$ or, respectively, truncation of $g_n$ in response to low estimates amplifying any existing bias. This last bias is estimator-specific, as it is only relevant for estimators which are using $g_n$. However, this includes all state-of-the-art doubly robust methods. Doubly robust methods are effectively reduced to g-computation when the last source of bias is too high \cite{petersen2012diagnosing}. But, while this means one cannot reap the benefits of this model class, if this is the only problematic source of bias, it is not relevant for estimability as $\bar{Q}_n$ is consistent and does not suffer from sparsity issues. However, as both sparsity and consistency are not black-and-white categories in practice, it is to be expected that this last source of bias will slightly increase bias for a doubly robust estimator. This holds, even if consistency and sparsity for the estimation of $\bar{Q}_n$ are satisfactory, compared to a world where the estimation for both $\bar{Q}_n$ and $g_n$ is satisfactory in this regard.
% g amplification is partially model-dependent, as it depends on the estimation of $f(a|l)$ and not all models even use this. Not the focus here. Unfortunately, extrapolation doesn't save us here. For this problem verweisen wir auf solutions die es schon gibt. this will often be covered by 1, but not always. If this turns out to be a problem, I guess we just have to rely on $Q$ and can't really benefit from $g$.

As the first and last source of bias are estimator-specific, meaning they can in theory be completely avoided by choosing a different estimator, the second and third source, which can be summarized as sparsity, are targeted in our diagnostic. The last source of bias is always targeted indirectly, but is addressed more explicitly in a second more complex variation of the diagnostic. It is important to note that not all of the 3 biases are caused by stochastic violations of positivity alone. But since all types of finite sample bias threaten the estimability of our estimand given the data set in the same way as finite sample bias due to stochastic positivity issues, i.\ e., due to the lack of data where it is needed, considering stochastic positivity violations together with other types of sparsity in our feasibility diagnostic is preferable to limiting this diagnostic to strictly positivity-induced issues.

\subsection{Extrapolation and Stochastic Positivity}\label{extra}
%*
 %[hier noch etwas darüber, dass das Konzept von stochastic positivity violations iwie nicht so viel Sinn macht, wenn man nicht Extrapolation berücksichtigt, aber ich glaube das hat noch nie jemand gesagt und ich weiß nicht ob ich das jetzt alleine so gut begründen kann?]
 %[und ich könnte hier noch was darüber schreiben, dass 100\%ige deterministic violations iwie sehr sehr selten sind, bzw nicht wirklich unterschiedlich zu stochastic gegeben extrapolation]
When predictions are made using continuous covariates, one typically has to resort to inter- or extrapolation. This can be seen as a sparsity problem as the number of values the continuous covariate can take is higher than what the finite data sample can provide. In this work we consider interpolation as a special case of extrapolation and therefore summarize both under the term extrapolation. As extrapolation is a necessity for the estimation of a causal effect, at least when the treatment or some covariates are continuous and, as discussed before, extrapolation is a possible solution to stochastic positivity violations, one always has to consider stochastic positivity in conjunction with extrapolation in the continuous treatment case.
In fact, the mere concept of a stochastic positivity violation only makes sense if extrapolation is considered, otherwise finite data on a continuous treatment would always imply a stochastic positivity violation almost everywhere.

As extrapolation is at the core of stochastic positivity in continuous treatment cases, the question remains how models extrapolate based on finite data. In a standard linear regression model, every data point potentially influences the estimated line in the same way. For more advanced state-of-the-art models like semi-parametric regression \cite{ruppert2003semiparametric} or other cross-validated loss-based learning, e.\,g.\ tree-based methods \cite{james2023tree}, estimation is typically done locally, meaning close data points influence the estimate at a certain point more than data points that are farther away. Therefore, when considering extrapolation, stochastic positivity violations are present in areas where there is not enough data in their surroundings. This leaves two questions open, which a diagnostic that identifies sparsity issues should deal with: 
\begin{enumerate}
    \item What is \textit{close enough} to be considered surrounding? 
    \item What is \textit{enough data}?
\end{enumerate}

The diagnostic we built based on these two questions can be used in two different ways. Firstly, it can be used in a data-centric way, meaning the focus lies on identifying sparse regions in the data, which in turn will influence estimation and lead to bias, but the focus is not yet on the exact consequences. Secondly, it can be used in an estimator-focused way. Here, the diagnostic specifically shows which parts of the estimation face sparsity issues.

%Extrapolation is necessary anyway cause continuous treatment, extrapolation can smooth over positivity problems to a certain degree and extrapolation is typically done locally, i.e. the estimator learns from close points (semi-parametric regression, cross-validated loss-based learning). The more extrapolation, the more model misspecification.
%The diagnostic is based on the idea that in flexible regression approaches (ML, semiparametric regression) we predict for a certain point given information provided by close neighbors to that point.

\subsection{The Diagnostic in Simple Terms: A Data-centric Diagnostic}\label{simplediasec}

At its core, the diagnostic we propose is a rather intuitive tool. It breaks down the $p$-dimensional information of whether there is data close to the points for which we want to get a prediction into one dimension by using a proximity measure. In the construction of this proximity measure the leading question has to be ``How far are we willing to extrapolate information for this specific estimand?'' which corresponds to the first question of the last subsection: ``How do we define \textit{close enough}?''. If we are interested in the effect of height on job interview success it makes sense to assume that in order to predict the success of a person of 1.80\,m the outcome of a person of 1.81\,m is very helpful. If the question concerns whether a person can pass trough a door of 1.80\,m without bending down, extrapolation is less successful. While in theory \textit{close enough} should be decided on using domain knowledge, we realize this is rarely feasible for practitioners. If domain knowledge is not readily available, it might make sense to come up with a best-case, worst-case and median-case scenario. Later on, we also provide some rules of thumb.

Whether a point is close or not should ideally not be a black-and-white issue. Therefore, we propose the use of a ``kernel of information'' $k$, which weights points according to the distance of the point we want to make a prediction for, $o_i^{int}$, to all values of $o^{obs}$, i.\ e., to all observed data.
The assumption is that the more distant an observation $o_j^{obs}$ is from $o_i^{int}$ the lower is the information contained within it to help make a prediction for $o_i^{int}$.
The kernel $k$ is most easily formulated as taking the distance between $o_i^{int}$ and $o_j^{obs}$, as the kernel needs to be centered around $o_i^{int}$, and therefore should fulfill the 
following properties:

\begin{description}
    \item[Property 1:] $k(0)=k(o_i^{int}-o_i^{int})=1$
    \item[Property 2:] $ k(o_j^{obs}-o_i^{int}) \subseteq [0,1]  \,\,\forall i,j \in 1,...,n$
\end{description}

The diagnostic measures the effective data points (EDP), where points further away from the point of interest are discounted.

\begin{figure}[t!]
    \centering
    \begin{subfigure}[t]{0.4\textwidth}
        \centering
        \includegraphics[width=1\textwidth]{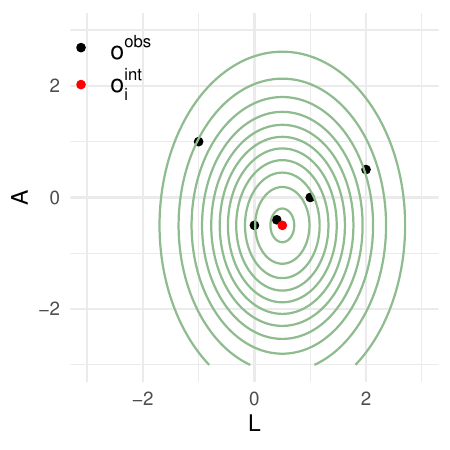}
        \caption{Example of a kernel $k(x)=k(x-o_i^{int})$ for a given $o_i^{int}$: The values the green kernel has at each data point are summed up and represent the number of effective data points for estimation of outcome $y_i^{int}$.}\label{kernelpic}
    \end{subfigure}%
    \hspace{0.07\textwidth}
    ~ 
    \begin{subfigure}[t]{0.4\textwidth}
        \centering
        \includegraphics[width=0.97\textwidth]{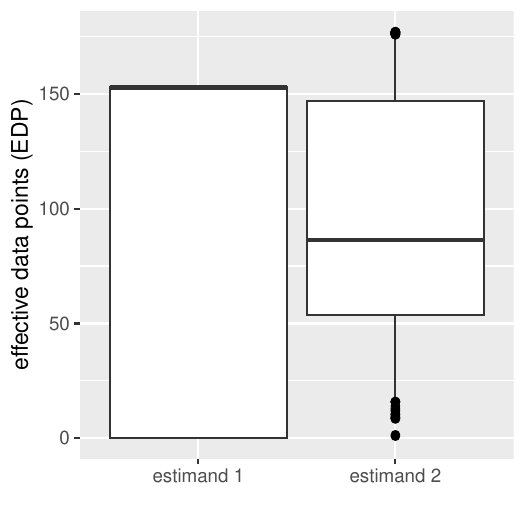}
        \caption{Example comparing two estimands using the data-centric diagnostic: For every observation the number of effective data points (EDP) is shown. EDP close to zero represent observations for which an estimation is difficult. Here, estimand 2 is clearly the superior choice.}\label{simpledia}
    \end{subfigure}

    \caption{A fictional example illustrating the kernel and the data-centric diagnostic}
\end{figure}

Figure \ref{kernelpic} shows an example of a Gaussian kernel used in a simple setting with one confounder $L$ and one treatment $A$. The Gaussian lends itself well to our diagnostic, as it decreases monotonically from the center which makes sense for a proximity measure. While many other kernel shapes are theoretically possible, we focus on the Gaussian in this paper as there is no obvious disadvantage and it simplifies the analyses.
% maybe a bit on semiparametric models here?
The steepness of the kernel should be decided on by how far one is willing to extrapolate in each direction of the data. If the relationship in one direction is known to be close to linear/very smooth, then the kernel should be chosen as less steep. Effectively, we are weighting down data points based on how much we think we can learn from them.
The pseudo-code for the calculation of the diagnostic is given by Algorithm \ref{diagnosticpseudocode}.

\begin{algorithm}[b]
\caption{The kernel-based sparsity diagnostic}\label{diagnosticpseudocode}
\begin{algorithmic}
    \State \textbf{Input:} $n$ values for $o^{obs}$, $n$ values for $o^{int}$, and diagnostic kernel $k$
    \State \textbf{Output:} $n$ values of ``effective data points''
    \ForEach {$o_i^{int}$, with $i=1,\dots,n$}
        \State $\text{EDP} = \sum_{j=1}^n\left(k(o_j^{obs}-o_i^{int})\right)$
    \EndFor
\end{algorithmic}
\end{algorithm}

After the new value $o_i^{int}$ is calculated based on $o_j^{obs}$ and the intervention scheme for $j=0,...,n$, we iterate over the values $o_i^{int}$ for $i=0,...,n$. For each value of $o_i^{int}$, the kernel function is evaluated on all $o_j^{obs}$ and the results are summed up. 
The $n$ values, one for all values of $o^{int}$, are then returned and can be displayed as a boxplot. This boxplot can then be used in order to answer the second question posed in Section \ref{extra} in a more nuanced manner. 
The lower ends of the boxplot, i.\ e., the points that are not well supported in the data, potentially indicate a problem and are therefore more interesting. Therefore, we modified the typical boxplot to end the whiskers always at the 5th and 95th percentile for this work. In order to illustrate how the diagnostic can be used, we provide a (fictional) example in Figure \ref{simpledia}. In this plot, two estimands are compared based on the diagnostic values for each data point. The plot shows that estimand 1 has more than 25 \% of the data completely unsupported, meaning these subjects received a treatment intervention such that they are now in areas of the $p$-dimensional space where no data were observed close by. For estimand 2, most intervened on data points are in areas of high support, however, there are a few data points with support close to zero. In this case, estimand 1 is expected to have a high bias due to sparsity. Estimand 2 will be biased because estimation is bound to perform poorly for a few badly supported data points, however, these data points can be examined and how much impact they might have can be gauged based on domain knowledge. Alternatively, these data points could be excluded from the study via a change in estimand by adjusting the target population.

\subsection{A More Thorough Look: An Estimation-focused Diagnostic}

This simple version of the diagnostic ignores that in the estimation of causal effects, there is not one model that can potentially be estimated, but two: $\bar{Q}_n$ and $g_n$, as discussed in Section \ref{sourcesofbias}. 
In order to consider how sparsity influences estimation, we have to examine the two models $\bar{Q}_n$ and $g_n$, separately. Estimation of $\bar{Q}_n$ and $g_n$ relies on having observed enough data points $o^{obs}$ close to each value that has to be predicted, i.\,e.\ the intervened on observations $o^{int}$. Specifically, for $\bar{Q}_n$ that concerns data dimensions $(A,L)$ and for $g_n$ only dimensions $L$. If the meaning of \textit{close} coincides for the two models, i.\ e., extrapolation can be done similarly for both models, then sparsity for $g_n$ implies sparsity for $\bar{Q}_n$. As was noted before, if $g_n$ is used in estimation, it is not only bias in $g_n$ that poses a risk for bias in the overall estimate. If estimated values of $g_n$ are small they can amplify bias since the inverse of $g_n$ is used as a weight. Therefore, if we want to provide a thorough look at estimability with our diagnostic, the size of the weights $g_n$ might produce for estimation needs to be considered as well.

For the estimation-focused diagnostic, the simple data-based diagnostic has to be applied to $(A,L)$ with a kernel describing extrapolation for $\bar{Q}_n$, and to $L$, with a kernel describing extrapolation for $g_n$. In addition, the diagnostic can be used in order to get a rough idea on the values for the weights $w_i=1/g_n(o_i^{obs})$, as the weights are based on $g_n$. Checking the weights for extreme values is recommended in literature in order to screen for positivity issues, specifically bias number 4, but as Petersen et al.\ \cite{petersen2012diagnosing} point out, only considering the weights which are to be used in the model is not enough as there could be regions which do not contain any data points. Some of these data points could, however, have a high weight had they been included in the estimation and been an indication of a sparsity issue. Therefore, we recommend checking the weights of \textit{ideal} data points, which lie exactly at $o_i^{int}$, resulting in $w_i^{ideal}=1/g_n(o_i^{int})$. Those weights can be roughly gauged using a kernel with the idea of counting effective data points at $o_i^{int}$. If there are very few observed data points, the inverse of the amount of observed data points is going to become very large, which is an indicator for sparsity inflating the bias.
We suggest using a rather steep kernel as the meaning of the kernel shifts from extrapolation of information from other points to knowledge of the data distribution at this exact point. Instead of the ability to extrapolate, the kernel shape now describes the smoothness of the estimated density.

\begin{wrapfigure}{r}{0.4\textwidth}
  \begin{center}
    \includegraphics[width=0.38\textwidth]{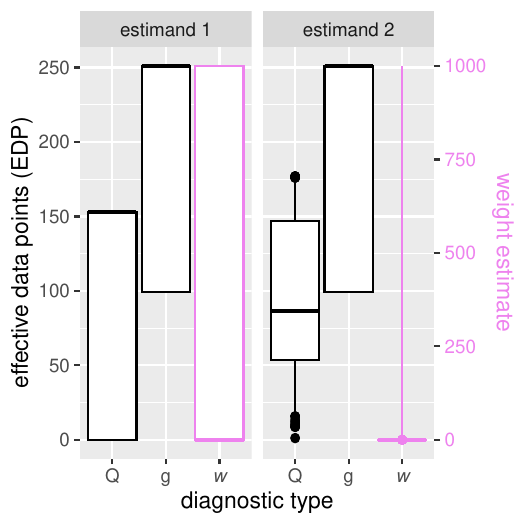}
  \end{center}
  \caption{A fictional example illustrating how the estimator-focused diagnostic can help decide between estimands.}\label{extendeddia}
\end{wrapfigure}
A diagnostic with these three subdiagnostics could look like the fictional Figure \ref{extendeddia}. Here, estimand 1 indicates severe problems for the  estimation of $\bar{Q}_n$ and while $g_n$ seems to be well estimable from the data, the weights for a lot of data points are extremely high (we chose to set a maximum weight of 1000 for the graphic, so the actual weight is even higher for a lot of points). Therefore, neither g-computation nor IPTW look promising here. For estimand 2, $g_n$ is well estimable again, but there are still over 5 \% of weights over 1000, so IPTW is not recommended here either. G-computation, however, might be a more promising route to take. While a few points show low data support, most data points are well supported. It might be worthwhile investigating why some points have so little support and whether a slight change of estimand 2, e.\,g., by excluding some observations from the analysis known as restricting the population (more on that in Section \ref{realdata_stp1}, where we perform a similar adjustment), might be even more estimable, but even if that is not possible, g-computation  does appear to be a promising approach for estimation based on the EDP.

The estimator-focused diagnostic is more meaningful than the data-centric diagnostic, as it provides the applied researcher with an exact understanding of where each model faces issues in estimation. However,
it is also more complex and requires a very thorough understanding of how estimation is supposed to behave. Therefore, we expect the data-centric approach to be more relevant to applied researchers.

\subsection{Kernel Choice and Hyperparameters}

%C2 other options for standard variation
%C3 (sim)

%c3 ordinal/cat variables
%c5 nadaraya-watson
%c7 kernel, h->0, and h->1
%c9 uniform kernel

In this section we provide some guidance on the choice of kernel, discuss how categorical variables can be dealt with, and present some rules of thumb on how to decide on hyperparameters.

The kernel represents how information is localized in the data and should be chosen in accordance with how extrapolation is expected to work in the data. Therefore, it is related to the model choice, which has to be based on similar considerations. For example, if the true function that is to be estimated is rather smooth, the kernel can be wide and a potential Generalized Additive Model estimator can use fewer basis functions. Many estimators use kernel approaches to assign more weight to data points that are close to the point of interest, leveraging the simple idea that nearby observations are typically more informative than distant ones. Kernel regression for example may be based on $\widehat{\mathrm{E}}(y|x)=\sum_{i=1}^n\frac{K_h(x_i-x)y_i}{\sum_{i=1}^n K_h(x_i-x)}$, with kernel function $K(t)=\frac{1}{h} K(\frac{t}{h})$ where $h$ is the bandwidth, in order to model a relationship of independent variable $x$ to dependent variable $y$ in a data set with sample size $n$ \cite{liu2014feature}. Here, the quantity $nh$ is often referred to as the \textit{effective sample size} as it describes the amount of data which is locally used for estimation \cite{chen1996empirical}. This quantity is rather similar to the idea of effective data points presented in this paper, as it is a local measure of information in the data. While the effective sample size refers to the information used in estimation, our effective data points aim at gauging the amount of information that is present in the data, which could theoretically be used by an estimator. The bandwidth of the kernel $h$ is only a descriptive measure of how extrapolation is conducted in the estimator, whereas the diagnostic hyperparameters, which control the steepness of the kernel, aim to describe how extrapolation should ideally be performed.

In the simplest case, the kernel function could be a uniform kernel $$
k(o_j^{\text{obs}}-o_i^{int}) = 
\begin{cases}
1, & \text{if }  \| o_{pj}^{obs}-o_{pi}^{int} \| \le h_p \,\forall p\\
0, & \text{otherwise,}
\end{cases}
$$ where $p=1,...,P$ are the dimensions of the data. This  reflects a binary choice sorting data points into the categories of \textit{useful for estimation}, which corresponds to assigning a value of 1, or \textit{not useful for estimation}, assigning a value of 0. The hyperparameters $h_p$ in this case decide where the cutoff between those categories lies. This results in a highly interpretable diagnostic, which shows for each observation how many data points are within that predefined range.
Regardless of the kernel in use, there are two special cases of hyperparameter choices which can be expressed in accordance with the formula stated above as $h_p=0$ and $h_p\to \infty$ for all $p$. If $h_p=0$, the assumption is that one can only learn from data points that are exactly alike. Therefore EDP will show how many data points there are with the exact same values in all covariates. If $h_p\to\infty$, the assumption is that all data points carry the same amount of information independent of any (dis-)similarities. This assumption is useful for example in a linear case. If the true shape of the function to fit is linear -- and as a consequence a linear estimator is used in order to capture the truth as well as possible -- the similarity aspect becomes unimportant and all data points should be counted with a weight of one. However, this will lead to $EDP=n$, which is not a useful result. In this case, the diagnostic becomes obsolete.

In causal inference, the treatment and confounders taken together are typically not \ one-dimensional. Therefore a multivariate kernel is needed. A simple choice would be the Multivariate Gaussian, but other kernel types could be used as well. In particular, univariate kernels could be used to construct multivariate ones, which can be done, for example, using a product kernel or a radial basis function. For a product kernel the $P$ univariate kernels $k_p$, one for each dimension of $x$, can be used to create $k(x)=\prod_{p=1}^Pk_p(x_p)$. This is especially helpful if different types of kernels should be used for different variables, but also comes with the drawback that a specific kernel has to be designed fo every variable. The alternative, the radial basis function $k(x)=k(\|x\|)$, with distance measure $\|.\|$, may be simpler to define. For Gaussian kernels, the two methods are equivalent.

The diagnostic was developed especially with continuous variables in mind. However, it can be used with categorical variables as well. Often, categorical variables will be dummy coded and therefore binary. As a consequence, the kernel (only considering this single dimension) should return $1$ if the data points fall in the same category and can be set to some fixed value $c$ if they fall into different categories. Typically, we would recommend $c=0$ as it is unclear how this categorical feature will influence the outcome in connection with the other variables. However, in cases where the influence of the category is expected to be of lesser importance, a value $>0$ could be chosen, but the resulting EDP will have to be more carefully considered. For example, in a data set with the binary category of sex and using $c=0.5$, an EDP of 20, which might be considered satisfactory depending on the specific context, this could mean that there are 40 similar men but not one woman even though the intention is to  predict for a woman in this hypothetical.
% here we should maybe explain the single dimension diagnostic better

The considerations above highlight how difficult the choice of the kernel function is. Ideally, the kernel and its hyperparameters are chosen using expert knowledge on the true shape of the function to be estimated. However, we recognize that this knowledge is hard to come by and we provide some guidance in the following. With regard to the kernel choice, we recommend a Gaussian kernel is used as it is simple, familiar to most researchers, and fulfills the criteria for the kernel function stated above. This leaves the choice of the  hyperparameters, which control how steep the kernel is and therefore how far one expects to be able to extrapolate. First, we recommend thinking about the steepness -- the variance for the Gaussian case  -- of the kernel in terms of halves. How much would a data point need to differ from $o_i^{int}$ in only one dimension, meaning all other features are the same, for the point to contain half as much information as a point which exactly equals $o_i^{int}$. Once this question is answered for all dimensions, the kernel hyperparameters can be set accordingly. Making these kinds of assumptions might seem very odd, but we would argue that researchers already make a similar assumption. Whenever a researcher decides to model a curve with $P$ dimensions and $n$ data points, an assumption on the smoothness of the curve and therefore implicitly on how far extrapolation can work is made. While the diagnostic forces the applied researcher to make this assumption very explicit and provide hard numbers, the assumption is the same at its core. Nevertheless, we recognize that deciding on exact numbers for the hyperparameters is quite challenging even for a researcher well versed in the field in question. Therefore, we provide some rules of thumb based on the observed data. We recommend one standard deviation as half distance for variables in the adjustment set and half a standard deviation for the treatment effect, as this is the most important feature in the prediction model. Those are the values we used in the practical application in Section \ref{actualdata} and while the simulation in Section \ref{chap_simulation} does not use exactly these values (as one fixed hyperparameter was chosen for several data distributions in order to simplify interpretation), the values used here are similar. In both cases the rule of thumb provided satisfactory results. In case more expert knowledge is available, one could start from these basic rules of thumb and tweak the values based on this knowledge. In addition, one could formulate worst and best case scenarios in order to gain more insight into how dependent on these hyperparameters the results are. We recommend to halve the rule of thumb values provided for the worst case scenario and double them for the best case scenario. Rather than basing the hyperparameters on the standard deviation, which is not robust to outliers, one could base this choice on the median absolute deviation or the interquartile range. Alternatively, one could use the average pairwise distance of two points, which is harder to compute, but provides more information on distances in the data set which are at the core of the diagnostic and therefore might be more interpretable.

In very high-dimensional cases it might be impossible to find data points which are close to $o_i^{int}$ in every dimension. This is how the curse of dimensionality is reflected in the EDP. We provide a simulation study illustrating how the diagnostic behaves in higher dimensions as well as some considerations on how to combat these issues in Appendix \ref{sim2}.

%C2 other options for standard variation
%C3 (sim)

%c3 ordinal/cat variables
%c7 kernel, h->0, and h->1
%c9 uniform kernel

\section{Simulation Study}\label{chap_simulation}

As previously stated, publications in the field of causal research are rather focused on static interventions. Other intervention schemes and MTPs are much less represented in research. Therefore, MTPs are not well established and generally not well understood by practitioners. However, the use of MTPs is often recommended as a response to positivity issues, as they ``have the advantage that they
can be designed to satisfy the positivity assumption required for causal inference'' \cite[p.\ 846]{diaz2023nonparametric}. In the following simulation study, we provide insights into the behaviour of some of the most popular MTPs in different sparsity situations. We show that it is not obvious when stochastic positivity issues cause bias given an estimand and demonstrate that our diagnostic bridges that gap, providing the information needed to decide which estimands are feasible and which are likely to be heavily biased due to sparsity issues. The code for this simulation study is made available in full on Github: \href{https://github.com/KatyFisch/kernelbasedSparsityDiagnostic}{https://github.com/KatyFisch/kernelbasedSparsityDiagnostic}.

\subsection{Methods}

The simulation is kept as simple as possible in order to illustrate the issues MTPs face in light of positivity violations and how the diagnostic can help to understand the problem. It is meant as a proof of concept. An analysis of a more complex scenario is presented in the form of an applied study in Section \ref{actualdata}.
The data are simulated using the simple structural causal model $O = (L,A,Y) \sim P_0$, where $L$ is a binary confounder, $A$ the continuous treatment variable, and $Y$ the continuous outcome:
\begin{align*}
L &\sim \mathrm{Bernoulli}(0.3) \\
A &\sim \mathrm{Normal}(g_A(L),\sigma_A^2) \\
Y &\sim\mathrm{Normal}(0.2 \cdot L + g_Y(A), 0.05).
\end{align*}

%3 positivity levels: $A$ is drawn from three different distributions given $L$ with different levels of positivity violation
The function $g_A(.)$ and variable $\sigma_a^2$ control the distribution of $A$ and can be used to induce problems regarding stochastic positivity by rendering regions empty of either any observations or specific observations for a given value of $L$.
Three different pairs of $g_A(.)$ and $\sigma_a^2$ were examined:
\begin{itemize}
    \item $g_A(L) = 0.45+0.21L$ and $\sigma_A^2=0.14$ inducing a large overlap of observations for different levels in $L$ \textit{(overlapping data)},
    \item $g_A(L) = 0.26 + 0.55L$ and $\sigma_A^2=0.08$ inducing a minimal overlap of observations for different levels in $L$ \textit{(adjacent data)},
    \item $g_A(L) = 0.18 + 0.64L$ and $\sigma_A^2=0.05$ inducing no overlap (but a gap) of observations for different levels in $L$  \textit{(divided data)}.
\end{itemize}
% 9 cdrc: linear, 2 piece-wise linear, 3 step function, a sine wave, a polynomial and an edge-flattened polynomial
For $g_Y(.)$, which controls the shape of the causal dose-response curve (CDRC), a polynomial was used in order to provide a certain kind of complexity for the models to learn. As MTPs shift data points in the direction of treatment, the behaviour of this polynomial outside of the plotted interval of interest $[0,1]$ is relevant. In order to make the behaviour out of bounds simpler, the polynomial turns to a linear on the sides. The resulting data distribution and underlying truth, i.\ e., what the expected outcome would be if all observations received that treatment value, can be seen in Figure \ref{results:cdrc}. The sample size $n$ was set to 350. Two other functional shapes for $Y$ were investigated as well, namely a piece-wise linear shape and a step function. The specifications for both of these are given in Appendix \ref{addsim}. 

\begin{table}[t]
\centering
    %\begin{adjustbox}{angle=90}
    \setlength{\extrarowheight}{5pt}
    \begin{tabularx}{\textwidth}{L{2.5cm}|l|L{3.6cm}|L{4.5cm}}
         Intervention  & Formula & Dose $a_i^{int}$ & Example (medication) \\
         \hline\hline
         dose-response curve &  $\mathrm{E}\left(Y^{A=x}\right)$ & $x \,\, \forall i$ & giving everyone a dose of 5\,g \\
         naive shift-response curve  (SRC)&  $\mathrm{E}\left(Y^{A=a^{obs}+x}\right)$ & $a_i^{obs} + x \,\,\forall i$ & increasing everyone's dose by 5\,g\\
         stochastic SRC &  $\mathrm{E}\left(Y^{A\sim\mathrm{N}(a^{obs}+x, \sigma^2)}\right)$ & $a_i^{obs} + x \pm \epsilon_i \,\,\forall i$  & increasing everyone's dose by 5\,g \textit{on average}\\
         %dynamic SRC & $\mathrm{E}\left(Y_i^{A_i=a_i+x\mathbbm{1}_{u_i} }\right)$ &  $ \left\{\begin{smallmatrix}
         %   a_i + x   \hfill & \text{ if } 0<u_i := u(h_i) \hfill  \\
         %   & \\
         %   a_i \hfill & \text{ else} \hfill
        %\end{smallmatrix}\right.$ &  increasing only women's doses by 5 g (no change for men)\\
         %stochastic dynamic SRC & $\mathrm{E}\left(Y_i^{A_i\sim\mathrm{N}(a_i+x\mathbbm{1}_{v_i}, \sigma^2)}\right)$ &$ \left\{\begin{smallmatrix}
         %   a_i + x  \pm \epsilon_i \hfill & \text{ if } 0<v_i := v(h_i) \hfill  \\
         %   & \\
         %   a_i \pm \epsilon_i \hfill & \text{ else} \hfill
        %\end{smallmatrix}\right.$  & increasing only women's doses by 5 g \textit{on average} (no change for men) \\
         threshold response curve & $\mathrm{E}\left(Y^{A=\max(x, a^{obs})}\right)$ & $ \left\{\begin{smallmatrix}
            x \hfill & \text{ if } a_i^{obs} < x \hfill  \\
            & \\
            a_i^{obs} \hfill & \text{ else} \hfill
        \end{smallmatrix}\right.$  & giving a dose of 5\,g to subjects with a  dose lower than 5\,g (no change else) \\
         conditional shift response curve &  $\mathrm{E}\left(Y^{A=a^{obs}+x\mathbbm{1}_{a^{obs}+x\leq w} }\right)$ & $ \left\{\begin{smallmatrix}
            a_i^{obs} + x \hfill & \text{ if } a_i^{obs} + x \leq w_i\hfill  \\
            & \\
            a_i^{obs} \hfill & \text{ else} \hfill
        \end{smallmatrix}\right.$ & increasing the dose by 5 g, but only if this increased dose is not toxic (else no change) \\
        %problem: we are doing it in both directions at the moment
    \end{tabularx}
    %\end{adjustbox}
    \caption[Different intervention curves considered for the single time point simulation study]{List of the different intervention curves considered for the single time point simulation study. The function $w$ is user defined based on the subject's history $H$, which includes $A$. For the stochastic models, $\epsilon$ is drawn from $\mathrm{N}(\mu, \sigma^2)$, where $\mu$ is provided and $\sigma^2$ estimated using a linear regression model. }
    \label{stp:interventions}
\end{table}

Table \ref{stp:interventions} shows the different intervention schemes we investigated in this simulation study. As static schemes are the most prevalent in research, we use the corresponding CDRC as a benchmark to illustrate how a switch to an MTP changes both estimand and estimability. The causal parameter of interest is here  $\psi(P_0) = \mathrm{E}\left(Y^{A=x}\right)$. The next scheme we considered is a naive shift and its naive shift-response curve. Here, we estimate $\psi(P_0) = \mathrm{E}\left(Y^{A=a^{obs}+x}\right)$. It is the simplest type of MTP. For the naive shift, we also considered its stochastic variant, namely $\psi(P_0) =\mathrm{E}\left(Y^{A\sim\mathrm{N}(a^{obs}+x, \sigma^2)}\right)$, where the intervention scheme remains the same on average, but has a random component added. In addition to these rather simple MTPs, we examined a threshold MTP and a shift MTP (in order to differentiate this scheme from the naive shift, we will refer to this as a conditional shift MTP), which coincide with the two schemes highlighted by D\'{i}az et al.\ \cite{diaz2023nonparametric}. The threshold shift MTP only shifts observations below a certain threshold with estimate $\psi(P_0) =\mathrm{E} \left(Y^{A =\max(x, a^{obs})}\right)$. The conditional shift MTP only shifts observations where a specific function is below a certain threshold, i.\ e., $\mathrm{E}\left(Y^{A=a^{obs}+x\mathbbm{1}_{a+x\leq w} }\right)$. For the function $w_i=w(a_i^{obs})$, we simply used the maximum and minimum values (depending on the direction of the shift) occurring in the data for each given $L$, meaning no observations will be shifted to have a treatment value which exceeds the treatment range of the observed data of their $L$ group. We estimated shift-response curves for both of these. As we do not believe stochastic versions of either of these to be useful, we did not include them in the study.

For each of the $3$ stochastic positivity situations, the dose-response curve is fitted using the CICI package with a g-formula based on a generalized GAM. Additionally, we use a doubly robust method based on Kennedy \cite{kennedy2017non}. All MTP models are fitted with the R package lmtp's Sequential Doubly Robust Estimator \cite{williams2023lmtp}, which is the authors' recommended method, use two folds for cross-fitting, and the following algorithms in its Superlearners: glm, gam, glmnet, and earth.
The hyperparameters of the diagnostic were set to 0.15 as half distance for $A$. As three different data distributions are shown in this section, rather than using a different one based on the standard deviation for each scenario, a unified value was chosen to make comparisons easier.

%\begin{figure}[t]
%    \centering
%    \includegraphics[width=0.91\textwidth]{figures/polynomial-linear_dia_drc_wellnodia.pdf}
%    \caption{Dose response curves with differing levels of sparsity}
%    \label{results:sim}
%\end{figure}

\subsection{Results}
In this section we show the response curves for the different intervention schemes and compare estimates to the true response. The plots show how the true curves and the estimated curves differ and how the diagnostic could help provide the information where bias is to be expected without needing access to the true response.

\subsubsection{Causal Dose Response Curve}

\begin{figure}[b!]
    \centering
    \begin{subfigure}[t]{\textwidth}
        \centering
    \includegraphics[width=\textwidth]{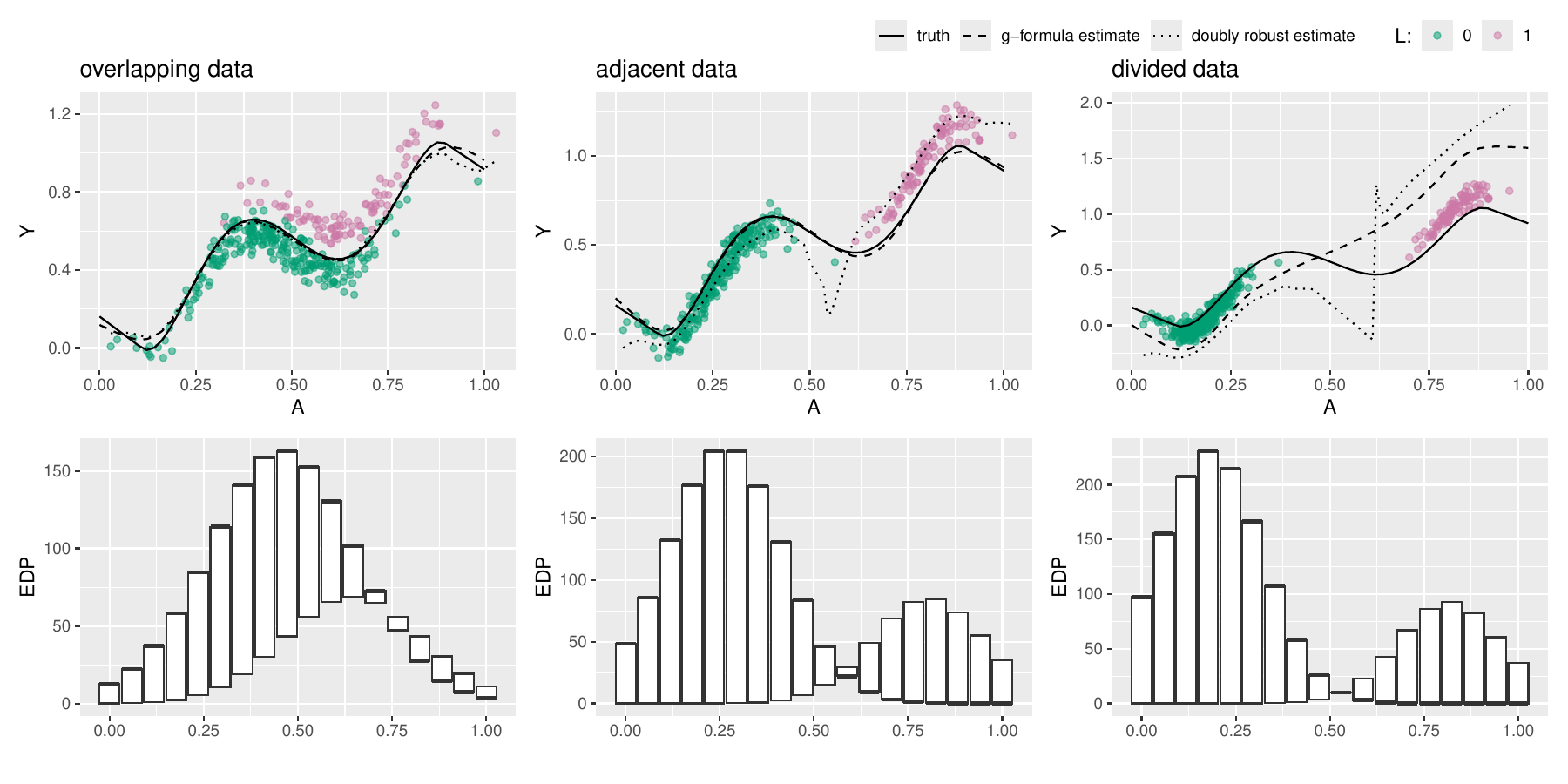}
    \caption{dose response curve}
    \label{results:cdrc}
    \end{subfigure}
    \begin{subfigure}[t]{\textwidth}
        \centering
\includegraphics[width=\textwidth]{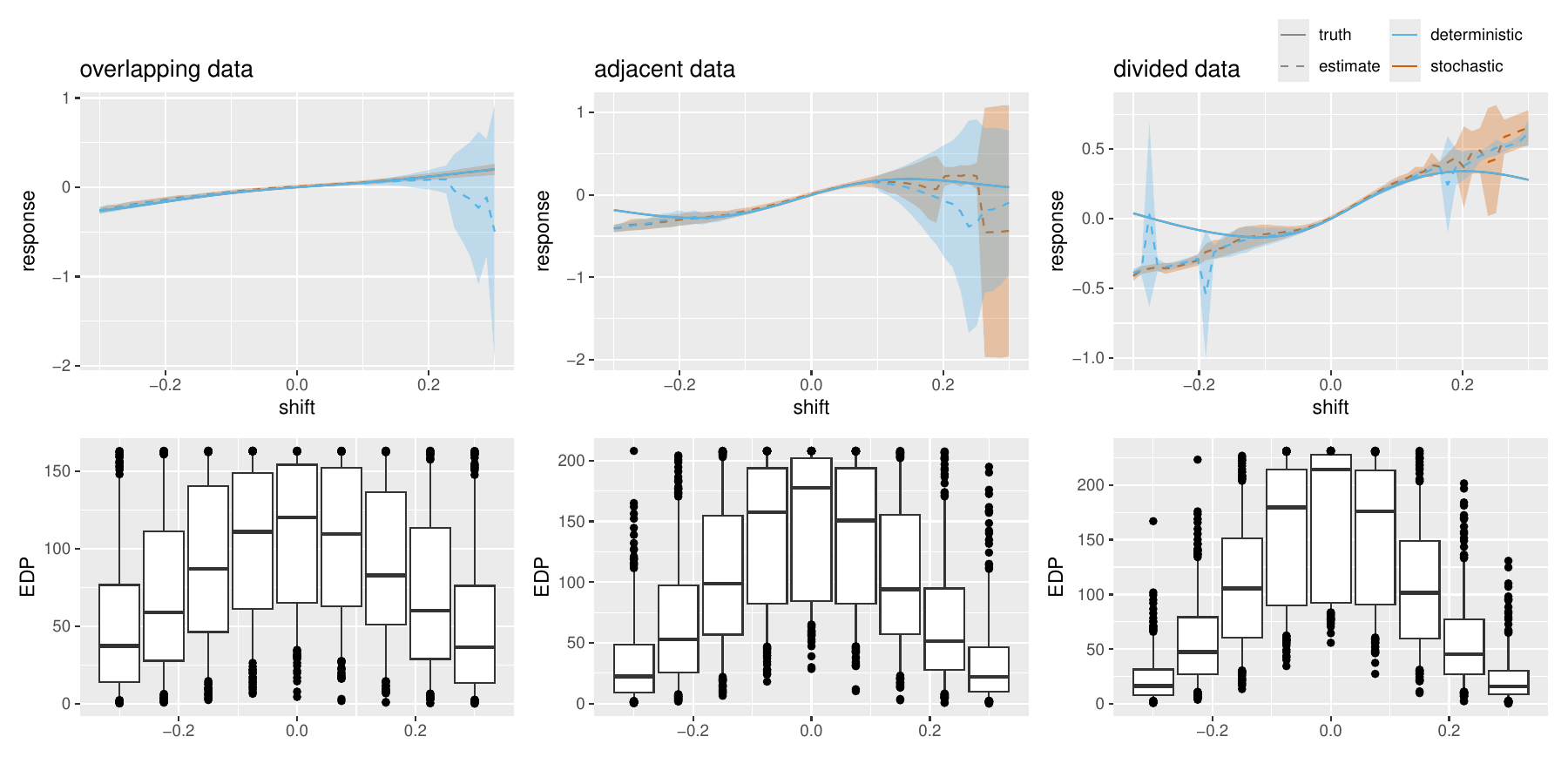}
\caption{naive shift response curve (MTP)}\label{results:src}
    \end{subfigure}

    \caption{Simulations for CDRC and naive shift response curve  with differing levels of sparsity}
\end{figure}

%todo: add confidence bounds

Figure \ref{results:cdrc} shows that in the overlapping data case there is a stretch of data with a lot of support for both $L=1$ and $L=0$, especially in the area where the density of $L=1$ is high. This is because $L=1$ is the rarer type of observation, meaning it represents a bottleneck. The functional shape of both the parametric g-formula estimate and the doubly robust estimate is very close to the truth. Both models are having trouble capturing the truth near the ends due to general sparsity. What is surprising is that around 0.25, where there is only data with $L=0$, both models still perform well. This is because the general trend is just continued from areas with more data. If this is not the case, for example if observations with $L=1$ had a value of 0.8 around this value, the estimates would be wrong. Basically, it is due to luck that extrapolation worked well here, although this was not to be expected. The diagnostic reflects a conservative estimate on where estimation works well. Therefore only areas with large support for both types of $L$ are shown as well supported, all other areas have lower ends of the boxplots near zero. An interesting side note is, that the boxplots only represent two values as in a fixed scheme all $a_i$ are set to the same value so the only difference in observations is in $L$. The median is therefore the effective support of all observations with $L=0$ and the other end of the boxplot represents the effective points for all observations with $L=1$.

For the adjacent data structure the g-formula estimate performs well, but the doubly robust estimate is off, especially in areas with little data. The diagnostic shows that there is no region with good data support for all observations, as we intentionally induced a stochastic positivity issue. The parametric model performs well because its parametric assumptions are in line with the truth, the doubly robust estimator on the other hand does not benefit from any parametric assumption and does not perform well. Again, the diagnostic is a careful indicator and does not rely on any parametric assumptions being correct, which is why it deems the whole range as poorly supported.

Lastly, both estimators are far from the truth for the divided data. The diagnostic shows a support of approximately zero for many data points for the whole range, correctly reflecting a data situation in which satisfactory estimation is not possible.

\subsubsection{Shift Response Curve}

In Figure \ref{results:src} the naive shift response curve is plotted. Due to the differing data distributions in the three simulations, it first has to be noted that the true curves are different in each scenario. With a naive shift model, some data points are always shifted outside of the range and the correctness of the result depends on extrapolation working well into these areas without data. In the overlapping data case, it is clear this extrapolation works better in the negative shift direction. The diagnostic, however, shows that there are problems with support in both directions. In the adjacent data case, and even more in the divided data scenario, the data points are more densely situated, therefore smaller shifts mean more data points leave the data range, which in turn means the estimation becomes more biased for smaller shifts. The more densely located data have another effect, too. In case of a shift of zero, the support is a lot better the more divided the data is.

%\subsubsection{Dynamic Shift Response Curve}

%\begin{figure}[t]
%\centering
%\includegraphics[width=\textwidth]{figures/polynomial-linear_dia_src_d.pdf}
%\caption{dynamic shift response curve (only $L=1$)}\label{results:dsrc}
%\end{figure}

\subsubsection{Threshold Response Curve}

\begin{figure}[t!]
    \centering
    \begin{subfigure}[t]{\textwidth}
        \centering
\includegraphics[width=\textwidth]{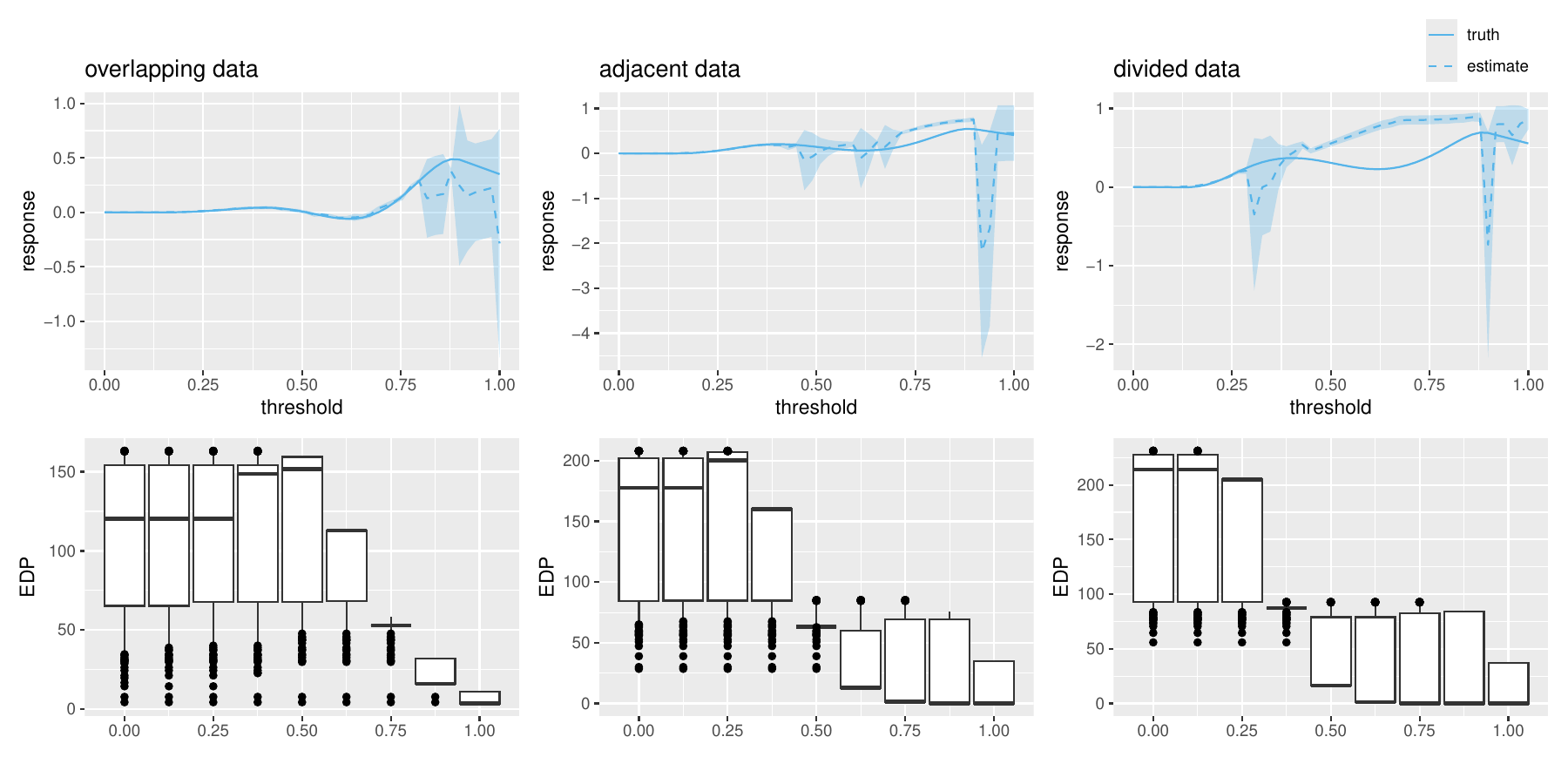}
\caption{threshold response curve}\label{results:trc}
    \end{subfigure}
    \begin{subfigure}[t]{\textwidth}
        \centering
\includegraphics[width=\textwidth]{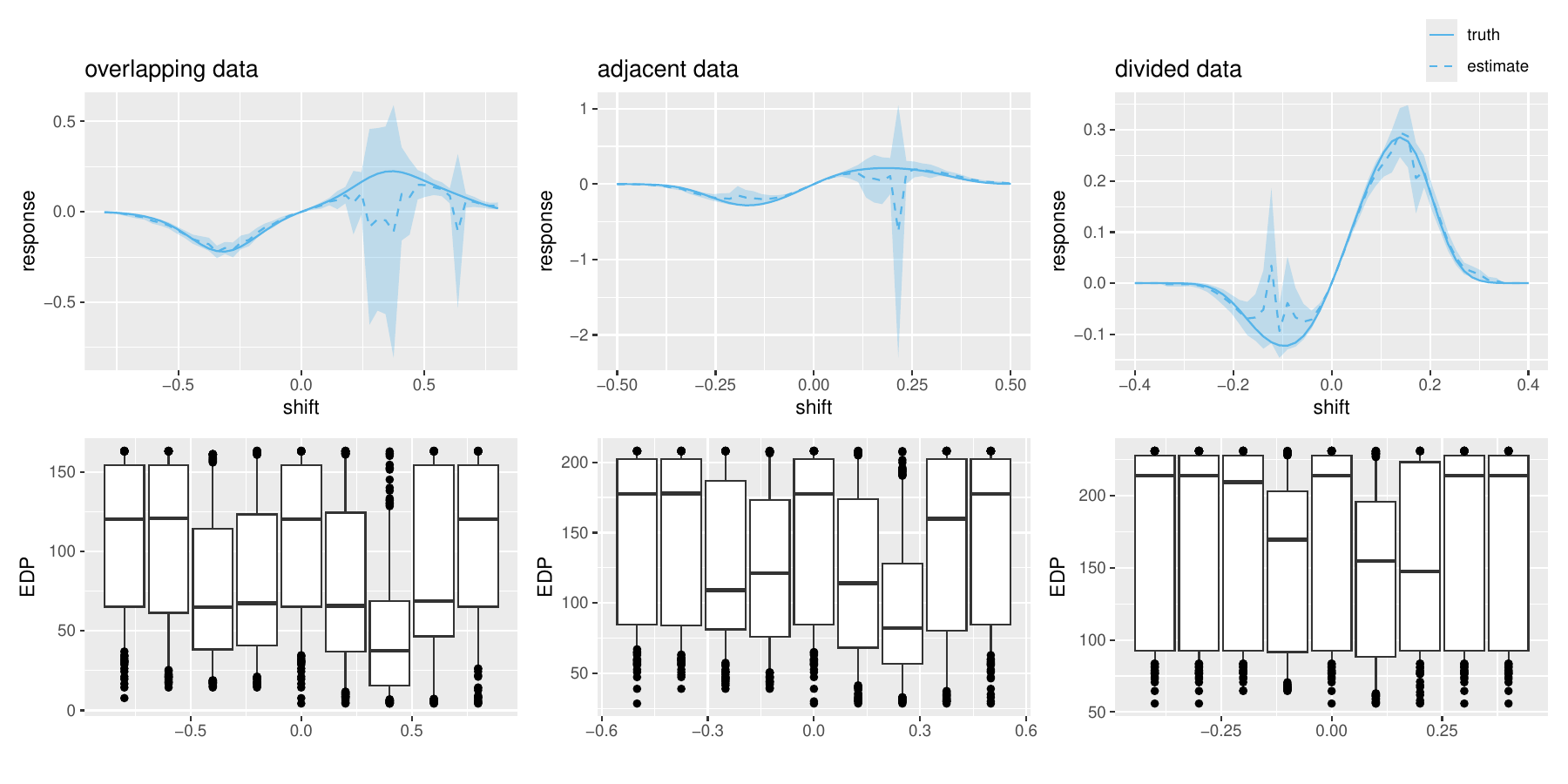}
\caption{conditional shift response curve}\label{results:csrc}
    \end{subfigure}

    \caption{Simulations for threshold and conditional shift response curve  with differing levels of sparsity}
\end{figure}

For the threshold response curve in Figure \ref{results:trc} the diagnostic shows that support becomes a problem as soon as the green data points are shifted outside of the area spanned by the green observed data points. For very low thresholds (almost) no observations are intervened on, therefore the result is the natural intervention where every subject receives their observed treatment. This area is smaller the more stochastic positivity is violated in this simulation. Therefore the model does not perform well for divided data, but is relatively stable for a large part of the threshold curve in case of the overlapping data. The diagnostic shows where the curve has support and can, without modeling, tell where the estimation will deviate from the true curve.

\subsubsection{Conditional Shift Response Curve}

Concerning the conditional shift response curve in Figure \ref{results:csrc}, observations are intervened on as long as they do not leave the observed range of their category, either $L=0$ or $L=1$. This means, positivity issues are largely avoided.
The diagnostic shows high support near zero and on the left and right side of the plots as there are no observations intervened on. The estimator does deviate from the true curve in all scenarios, however it does not leave the confidence bounds. The diagnostic shows high support for the whole curve especially in the divided data case. Therefore, this estimand is a viable option in all three scenarios.

\subsubsection{Further Simulations}

The results for the simulations of the piece-wise linear shape and the step function are shown in Appendix \ref{addsim}. These additional simulations especially highlight how outliers, for which a prediction cannot be estimated well, could potentially lead to a strong bias in effect estimation, if the true outcomes for these outliers are extreme. They also shine a light on true response curves of non-polynomial form and therefore demonstrate that the diagnostic works in various data scenarios.

\subsubsection{Conclusion}

The results show that stochastic positivity violations cause problems for MTPs whenever observations get shifted into areas where  extrapolation becomes unstable. This means it is vital to understand when a scheme shifts data into these areas. A diagnostic that shows how much data support there is in order to estimate the effect of the intervention scheme for every observation in the dataset can help get deeper insights.

\section{Practical Application}\label{actualdata}

In order to showcase how the diagnostic can be used by applied researchers, we apply it to data from a recent publication on determining targets for antiretroviral drug concentrations in children with HIV \cite{schomaker2024determining}. In this publication, the researchers evaluated the effect of a continuous exposure (drug concentration), measured over multiple time points, on a binary outcome (viral failure).  We show how the diagnostic can be used to identify problems with stochastic positivity and other kinds of sparsity, and develop suggestions to combat these. For the example at hand, this could help develop an estimand that leads to more stable estimates than those considered in the study. 

\subsection{Data and Research Question}

The data used for this study stems from the CHAPAS-3 (Children with HIV in Africa -- Pharmacokinetics and Adherence/Acceptability of Simple antiretroviral regimens) trial, a randomized controlled trial, where HIV-positive children between 1 month and 13 years of age in Zambia and Uganda were given different combination antiretroviral therapy regimens \cite{mulenga2016abacavir}. The current study only uses the subset of participants who received the drug efavirenz, which leads to a sample size of 125 children. The authors conducted a complete case analysis, which they had previously shown to be valid for the question explored below given the missingness mechanism they assumed \cite{holovchak2024recoverability}, leaving 58 children to be analyzed.

The target variable for this study was \textit{viral failure} $V\!F$, defined as $>100$ copies/mL. The exposure variable was not the amount of medication taken, but the so-called \textit{mid-dose concentration of efavirenz} $EFV$, meaning the concentration after 12 hours as medication is taken every 24 hours. This rather atypical exposure is utilized because the target concentration window is at the core of the research question. This concentration was not directly measured, but calculated from measurements taken 2-4 hours after intake through a population PK model \cite{bienczak2016impact}. The data is longitudinal in nature and follow-up was at $t=6,\ 36,\ 48,\ 60,\ 84, \text{ and } 96$ weeks. 
Viral load was measured at all time points except 6 and efavirenz concentrations were assessed all weeks except 48 and 96.
Therefore, the goal of the analysis is to determine $P(V\!F_t^{a})$ with $t=36,48,60,84,96$ and $a$ signifying a treatment scheme of efavirenz. This is in line with previous research on the same data set  \cite{bienczak2016plasma, holovchak2024recoverability, schomaker2024determining}. 
The sufficient adjustment set comprises only \textit{weight}, here in a logarithmized form, and a time-independent indicator of general \textit{adherence} to the treatment protocol \cite{schomaker2024causal}. 
%The children's weight was measured at every time point, including the beginning of the study. Adherence was measured using medication event monitoring systems, which measure when medication bottles are opened, however, the caps where not used as often as planned due to funding and practicality. Therefore, the variable adherence is used as a time-independent binary indicator, which summarizes the available information for each child in the study.
Lastly, the \textit{metabolization group} is of interest as a potential effect modifier as it is one of the main contributors to the high between-subject variability of efavirenz treatment \cite{bienczak2016impact}. 
%The variable categorizes all children into one of four groups from fast to very slow metabolizer according to two specific genetic markers, which influence how fast efavirenz is metabolized. 
The research question of the study 
concerns the CDRC for the antiretroviral drug efavirenz (more precisely in this case a Causal \textit{Concentration} Response Curve) with a focus on 
 the therapeutic window of the target concentration. In both adults and children the mid-dose target concentration window  is typically given as between 1 mg/L and 4 mg/L, but especially in children the validity of this window is partially unclear \cite{bienczak2016plasma, schomaker2024determining}. The upper end of the target window limits the probability of adverse events, but as these were rare in the trial, this upper limit cannot be evaluated well using the data at hand. Therefore, the focus lies on the lower bound, which limits the risk of viral failure. 
 %One counterfactual question of interest would therefore be ``if every child in the study had a mid-dose concentration of 1 mg/L, would viral failure be sufficiently small'', where a failure in 5 \% of the children is used as a threshold in the study. 

\subsection{Single Time Point Analysis}

At first, in order to simplify the problem at hand, the question is reduced to a single time point target. The variable set is therefore reduced to the initial weight, general adherence, and, if effect modification is of interest, the metabolization group. In addition, the mid-dose concentrations at weeks 6 and 36 are the exposure of interest, which we set to be the same at both time points in the intervention scheme for simplicity. The target is the viral failure measured at week 36 in the form of a dose-response curve. While $a$ can take higher values, we restrict the intervention to a medically more relevant window of $a^{int}\in[0,6]$ \cite{schomaker2024determining}. The estimand is therefore $P(V\!F_{t=36}^{EFV_{t=6}= EFV_{t=36}=a^{int}}=1)$ for $a^{int}\in[0,6]$, so the dose-response curve of viral failure in week 36 had the mid-dose concentration of efavirenz been $a^{int}$ in both weeks 6 and 36.
%One way to determine a target concentration window is to assess the CDRC, with a focus on where the curve is above 5 \%. 
In the following, the diagnostic is used in order to analyze how much stochastic positivity/sparsity issues influence the estimation of a CDRC. This is done first for a general causal estimate and then in a second analysis for a causal estimate which includes the estimation of effect modification by the metabolization group.

\subsubsection{Causal Analysis}\label{realdata_stp1}

Before the diagnostic can be used, the hyperparameters have to be fixed. We choose a Gaussian kernel like in the simulation study above. Next, the hyperparameters controlling how wide 
\begin{wrapfigure}{r}{0.6\textwidth}
    \centering
    \includegraphics[width=0.6\textwidth]{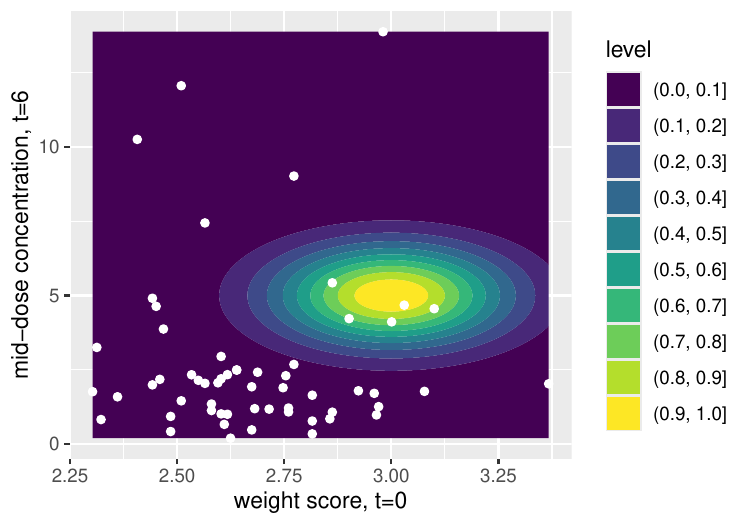}
    \caption{The kernel density with its chosen hyperparameters for an $o^{int}$, with weight 3 and mid-dose concentration 5 plotted against the observed data points}\label{dia_kernel}
\end{wrapfigure}
or narrow the kernel is in every direction need to be set.
Specifically, the question ``How far do we think we can extra- and interpolate in each dimension?'' has to be asked. 
For the binary variable adherence, we assume that no extrapolation is possible. For the other adjustment variable weight, we assume that an observation which shares every other variable, but deviates in weight by one standard deviation, should get a diagnostic value of 0.5, meaning it would be treated as half an observation of support. As mid-dose concentration is the exposure variable, we set a more narrow shape in that direction with only half a standard deviation leading to a half observation of support. We can illustrate how that kernel will influence the diagnostic by assuming we are interested in ascribing a concentration of 5 to a person with weight score 3, see Figure \ref{dia_kernel}. If we only consider these two dimensions, this leads to a diagnostic value of roughly 4.3, which is mostly due to the 5 points close by. Plots like this can also help researchers gain a deeper understanding on what certain hyperparameter values mean in terms of extrapolation, and can help choose them according to their domain knowledge.

\begin{figure}[t!]
    \centering
    \begin{subfigure}[t]{0.5\textwidth}
        \centering
        \includegraphics[width=\textwidth]{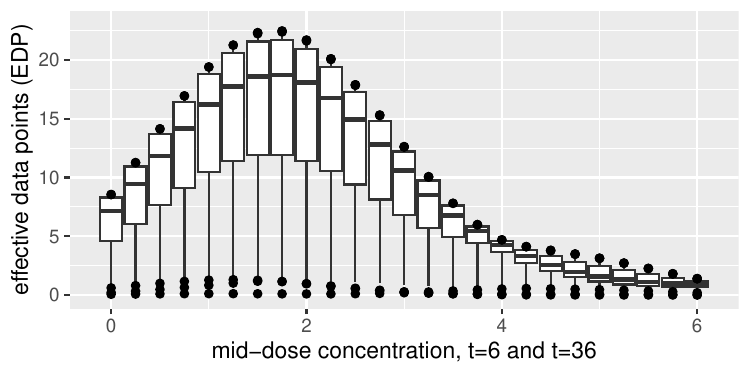}
    \caption{The diagnostic for the CDRC on the full data set}
    \label{causal_full}
    \end{subfigure}%
    ~ 
    \begin{subfigure}[t]{0.5\textwidth}
        \centering
        \includegraphics[width=\textwidth]{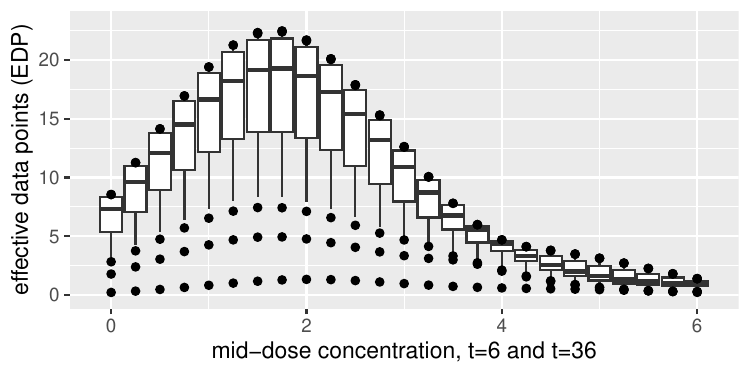}
    \caption{The diagnostic for the CDRC on the subset of adherent patients}\label{causal_adherent}
    \end{subfigure}

        \begin{subfigure}[t]{0.5\textwidth}
        \centering
         \includegraphics[width=\textwidth]{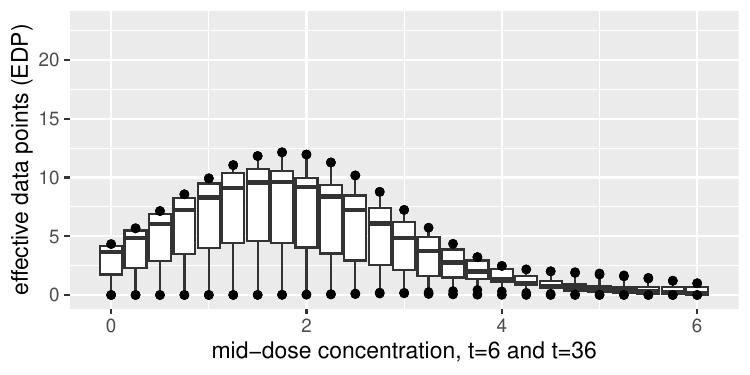}
    \caption{The diagnostic for a CDRC assessing effect modification by metabolization group}\label{effectm_single}
    \end{subfigure}%
    ~ 
    \begin{subfigure}[t]{0.5\textwidth}
        \centering
           \includegraphics[width=\textwidth]{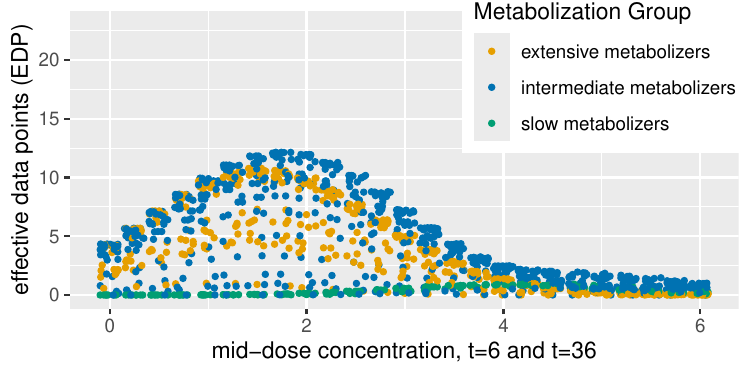}
    \caption{The diagnostic for a CDRC assessing effect modification by metabolization group (jittered and color-coded)}
    \label{effectm_bymetgroup}
    \end{subfigure}

    \caption{The diagnostic on different estimands for the CHAPAS-3 data}\label{realdata_stp}
\end{figure}

The diagnostic for the CDRC using this kernel is plotted in Figure \ref{causal_full}. For each mid-dose concentration in both week 6 and 36, the boxplots show  how much data support there is in order to calculate the effect of said concentrations on the sample. As the  lower limit of the currently used target concentration window is 1 mg/L, the CDRC around that value is especially interesting. We investigate all observations which had a low diagnostic score at concentration 1, as we expect a high bias for these predictions due to data sparsity. While it is difficult to decide what value of EDP is considered too low, only four observations have a score below 4 effective data points, all of them with a score $<1.4$ EDP. Therefore, these four points present clearly as outliers in the diagnostic. Considering the sample size was only 58, expecting an EDP$>4$ for all observations seems not feasible, while an EDP of around 1 is clearly not ideal. These considerations, while still subjective, made it rather clear which observations should be seen as potentially problematic for estimation and are therefore further investigated. Out of these four observations, three are non-adherent. As these are also the only observations who are classed as non-adherent in the complete cases sample, this is the main reason why the diagnostic score is so low for these individuals. 
This first sparsity issue identified cannot easily be combated by changing the treatment scheme, as the sparsity is induced not by the exposure variable, but a variable of the adjustment set. And since there are only three data points which share this value, not much can be done to improve the situation. A solution would be to collect more data, which is costly and not usually feasible, or to restrict $L$ by removing adherence from the adjustment set, resulting in a trade-off between bias due to sparsity and bias due to confounding. However, for this last option the magnitude of the induced bias is unclear. Another possible solution is to change the estimand by excluding non-adherent individuals. This leads to the changed diagnostic plot shown in Figure \ref{causal_adherent}. The three lowest scoring observations in this new  analysis are the three individuals with the highest weight, so if the data support shown by the diagnostic is still deemed unsatisfactory, individuals of higher weight could be excluded from the estimand in a further step by setting the target population to adherent observations with a weight below a certain threshold. While the diagnostics show that both of these changes in estimand improve data support, the clinical relevance of these estimands has to be judged by medical experts. In general, the diagnostic can help develop strategies to combat sparsity issues based on the data, but whether these strategies are actionable has to be evaluated by domain experts in a last step.

\subsubsection{Causal Analysis with effect modifiers}

In a second analysis, the causal question of interest is extended by including the metabolization group as an effect modifier of interest. We consider the whole complete cases sample and use the same settings for the kernel as previously. In addition, for the metabolization groups, we assume that no extrapolation is possible between groups. The diagnostic plot for the CDRC including the metabolization group is given in Figure \ref{effectm_single}. This shows a lot of observations have close to zero support. In order to gain insights into why that is, the same information is plotted again, but instead of boxplots a jittered scatterplot is created, where metabolization groups are shown by color, see Figure \ref{effectm_bymetgroup}.

This plot shows that slow metabolizers are the root of the sparsity problem. As their metabolism works more slowly than for other individuals, they tend to have rather high mid-dose concentrations and there is little support for them at the 1 mg/L mark. This is a typical example of a positivity violation in the exposure variable. Assuming that a slow metabolizer received an efavirenz treatment in accordance with the trial guidelines, which are in turn based on the WHO guidelines, it is not possible for their metabolism to reach a concentration that low after just 12 hours. Therefore, researchers asking the question of ``what if'' would make a mistake, as that is unrealistic to happen.
This problem can be combated by switching the exposure scheme, for example to a dynamic scheme, where slow metabolizers are set to a concentration of 4 mg/L and extensive and intermediate metabolizers to a concentration of 1 mg/L, or a modified treatment scheme, where, again, intermediate and extensive metabolizers receive a concentration of 1 mg/L, but slow metabolizers just receive their natural exposure. Alternatively, the population of interest can be changed by excluding slow metabolizers from the study. Medical experts again have to determine, which of these options would be clinically relevant. For more information on the clinical background see Schomaker et al.\ \cite{schomaker2024determining}.

\subsection{Outlook: Multiple Time Point Analysis}\label{outlook:longi}

The longitudinal case goes beyond the scope of this paper, but we want to present a short overview on how longitudinal data could be approached. For this analysis we used the CHAPAS-3 trial data again. 
For the longitudinal case we illustrate the estimator-focused diagnostic. The goal here is to estimate the CDRC for a continuous treatment measured over multiple time points. While doubly robust estimands, for which doubly robust estimators can be constructed, exist for this data scenario, like the LMTP \cite{diaz2023nonparametric}, it is not possible to construct one for the CDRC \cite{schomaker2024causal, RN967}. Therefore, only IPTW and the g-formula can be used here.
As IPTW is known to handle extrapolation badly and the sample size is small, we focus on the g-formula here. For the diagnostic hyperparameters, we keep using a Gaussian kernel and the half distance to one standard deviation for the adjustment set and half a standard deviation for the exposure variables for the estimation of the outcome. In this example, we are interested in the estimation of two static intervention schemes. Firstly, $\bar{A}=(1,1,1,1)$ and secondly $\bar{A}=(1,1,10,1)$, where $\bar{A}$ describes the exposure at each of the four time points. The goal is to get an indication of how spikes in the concentration of efavirenz might change the outcome. 

\begin{figure}[t!]
    \centering
    \begin{subfigure}[t]{0.5\textwidth}
        \centering
        \includegraphics[width=\textwidth]{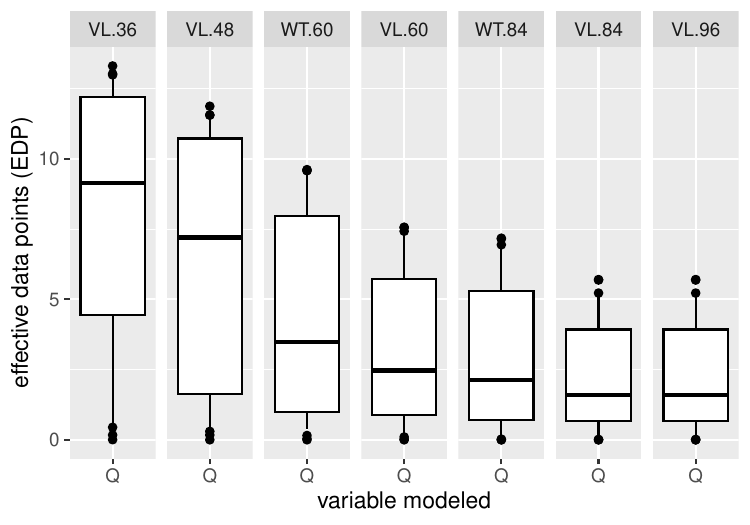}
    \caption{The diagnostic for intervention $\bar{A}=(1,1,1,1)$}
    \label{long_good_Q}
    \end{subfigure}%
    ~
    \begin{subfigure}[t]{0.5\textwidth}
        \centering
        \includegraphics[width=\textwidth]{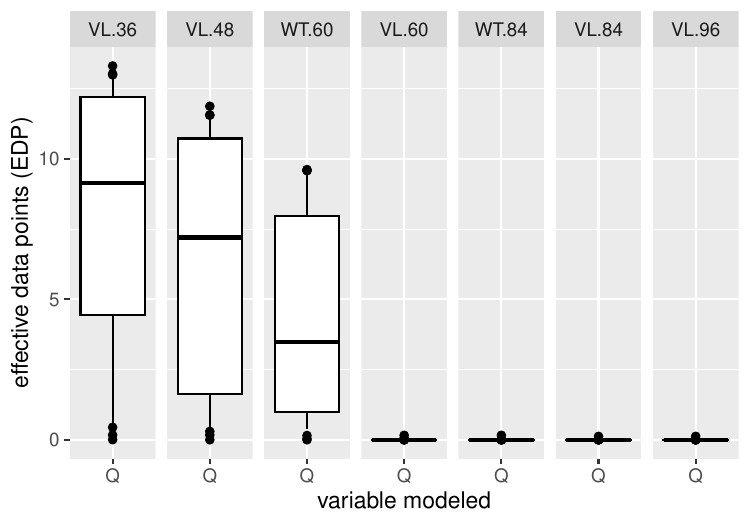}
    \caption{The diagnostic for intervention $\bar{A}=(1,1,10,1)$}\label{long_bad_Q}
    \end{subfigure}%
    \caption{longitudinal causal estimands}\label{real_long}
\end{figure}

For the parametric g-formula, $\bar{Q}_n$-models for all outcomes and post-intervention confounders need to be estimated. This results in seven models in total.
Figure \ref{real_long} shows the results of the diagnostic. For the first scheme, Figure \ref{long_good_Q} shows some issues with the estimation of each model in the parametric g-formula. While most observations show some support for this intervention scheme, a few observations are very close to zero and should be checked. Likely, those are the same points as the ones that caused issues in the single time point case. In conclusion, the models have good support for a large portion of the data, but there is room for improvement.
As for the treatment scheme with a spike, Figure \ref{long_bad_Q} shows that from the time of the spike on, there is almost no support and the predictions are likely to be heavily biased. Therefore, we conclude that it is not possible to estimate the spike model given the data at hand without risking a very large bias due to sparsity. The estimand should be changed, for example to a less drastic spike.

\section{Discussion}\label{discussion}

We developed a diagnostic for stochastic positivity issues and other types of sparsity and demonstrated its effectiveness in a simulation study as well as a practical data analysis.
We built the diagnostic such that it is first fixed how far extrapolation is possible and then the number of (effective) points for each observation is used to make an informed decision on the trade-off between sparsity induced bias and closeness to the estimand of interest.
In theory, the diagnostic can also be constructed the other way round, i.\,e., one first decides on how many data points are necessary for satisfactory estimation and then the diagnostic shows how far one has to extrapolate for each point in order to achieve that. However, we feel the approach we chose is more intuitive.

The clear advantage of the diagnostic is that it can be done as a first step before any modeling is necessary. Therefore, the diagnostic provides insights into which estimation efforts are promising and saves the researcher time by removing infeasible estimands from the to-do list before any effort is spent on them. One drawback, however, is the reliance of the diagnostic on hyperparameters, which should be based on domain knowledge. This knowledge is often imprecise or even non-existent which can greatly impact the performance of the diagnostic.

There are several possible extensions to this diagnostic measure. The choice of the kernel impacts the results, but it is unclear how to choose a kernel which describes well how all estimators included (for example in the Superlearner) extrapolate. We used a Gaussian kernel in this work, but the viability of other kernels should be explored. 
In addition, the added value of a data point is not yet considered. For example, it is intuitively clear that several data points which are similar in covariates provide less information on $o^{int}$ compared to the same amount of data points distributed in all directions around $o^{int}$. Therefore, the diagnostic could assume there is a lot of information at a certain point even though there is just a massive amount of information in one particular area close to it and no other data to guide in which way the estimated surface should bend in the direction of the point of interest.  There are some possible workarounds. For example one could use a Kullback-Leibler divergence between a density estimate of the weighted points and the kernel in order to gauge whether the points are well distributed within the kernel. However, this just adds to the complexity of the diagnostic and is still crude and not founded in any theory which would lead to an actual added value. More research is necessary here. Another possible extension concerns high-dimensional data. In such a case, the diagnostic might not be useful the way it is presented here, as no data points similar in all covariates can be found for a given observation. For example, there might be a very close patient sharing all 100 variables considered, except gender. The diagnostic would assume there is nothing to be learned from this patient. In order to combat this, the diagnostic can use a harmonic mean of the single dimension diagnostic values or set minimum values for every dimension of the data, so that a difference in gender will at most discount the diagnostic by, for example, 20\,\% (for more see Appendix \ref{sim2}).

One might wonder why the diagnostic is not based on a density estimation of $g_n$. Many methods already exist that use the density in order to decide on whether there is a stochastic positivity violation or not. However, we find this only makes sense in a discrete setting and, in addition, a question about sparsity is necessarily a question about data, not the data generating process behind it or the potential of data. Density estimation is highly difficult, and although one could build the diagnostic on a type of density estimation instead of pure data, but one would still have to deliberately design this to take the possibility of extrapolation into account. In addition, the estimate of the density has an uncertainty that the pure data does not have and we feel it would be less interpretable than the diagnostic we provide.

%Why not use density estimation? fundamentally it's not that different, but
%to answer the binary question is $f(x)=0$ I think it is difficult to tune density estimation in a way to answer that question well and including any domain knowledge is difficult. To do that well I still think you'd have to

A cautious researcher might wonder if choosing the estimand based on the data might bias the result as one of the so-called researcher degrees of freedom. However, as the diagnostic does only consider $A$ and $L$ and never uses $Y$, we do not think that this should be a concern.
Lastly, we want to mention, that, in principle, the diagnostic can be applied to any prediction problem and is not limited to causal inference.

\section{Conclusion}\label{conclusion}

Sparsity, especially in the form of stochastic positivity violations, is a big problem for causal effect estimation, but it is rarely discussed how to identify the problem. A commonly presented solution is to move the treatment scheme to a (longitudinal) modified treatment policy. However, MTPs are an extremely large class of estimands and it is not clear which MTPs will suffer from sparsity issues and to what extent.  In fact, as our simulation study highlighted, switching to an MTP is not a panacea and depending on the exact treatment scheme employed, different positivity assumptions need to be fulfilled.
The diagnostic we presented provides a simple way of diagnosing sparsity issues for a given estimand. It works for every kind of estimand, including static schemes, like a Causal Dose Response Curve, dynamic schemes, and Modified Treatment Policies, by comparing the observed data and the dataset under intervention. 
The question of sparsity is a $p$-dimensional problem and the diagnostic reduces this to a one-dimensional output, while also giving the researcher enough information to make the trade-off between estimation bias and closeness to the original research question.
In addition, the diagnostic is relatively simple and provides a lot of insight into estimability without having to estimate a single model.

\subsection*{Acknowledgements}
We are grateful for the support of the CHAPAS-3 trial team, their advise regarding the illustrative
data analysis and making their data available to us. 

We would like to particularly thank Diana Gibb,  Ann Sarah Walker, Iv{\'a}n D{\'\i}az, David Burger, Andrzej Bienczak, Elizabeth Kaudha, Paolo Denti and Helen McIlleron.

\subsection*{Funding information}
The research is supported by the German Research Foundations (DFG) Heisenberg Program (grants 465412241 and 465412441).

\subsection*{Author contribution}
All authors have accepted responsibility for the entire content of this manuscript and consented to its submission to the journal, reviewed all the results and approved the final version of the manuscript. 

\subsection*{Conflict of interest}
Authors state no conflict of interest.

\subsection*{Code}
We provide the full code to the simulation study presented in this work on Github: 

\href{https://github.com/KatyFisch/kernelbasedSparsityDiagnostic}{https://github.com/KatyFisch/kernelbasedSparsityDiagnostic}

\clearpage
\bibliography{cite}
\bibliographystyle{vancouver} 

\newpage
\begin{appendices}

\section{Positivity Assumption For LMTP}\label{extendedpositivity}

\subsection{Extending The Positivity Assumption To Non-Static Interventions}

%[todo: I should derive this formally, and the whole section for the longitudinal case]

The definition of the positivity assumption for a single time point for continuous treatments is given by 

\begin{center}
``if $f_L (l) \neq 0, f_{A|L} (a|l) > 0$ for all $a$ and $l$''
\end{center}

\cite[p. 255]{hernan2023causal}.
However, it is unclear what ``all $a$ and $l$'' are in this context, which is why we want to refine this definition here. ``All $l$'' refers to the adjustment set of the population considered in the research question. If the population only includes women, the positivity assumption does not need to hold for men. In cases where $l$ has more than one dimension, each dimension can be considered separately. For example, if age and height are included in $l$, there is no need to consider that babies will not be as tall as adults. The condition on the left hand side $f_L (l) \neq 0$ is not fulfilled for 190 cm tall newborns, therefore the right hand side does not need to be fulfilled either. The values for `` all $a$'' depend on the research question. In order to estimate the ATE for a given value of $a$, the positivity assumption only needs to be fulfilled for this one $a$. In order to estimate a causal dose response curve in the interval $\left[c,d\right]$, the assumption needs to be fulfilled for the whole interval. 

For static schemes, $a$ and $l$ can be considered separately as the same $a$ is applied to all observations. For dynamic schemes or modified treatment policies, $a$ and $l$ need to be considered together. The positivity assumption only needs to be fulfilled for pairs of $a$ and $l$ which are seen in the estimand, leading to
\begin{align}
f_L(l)\neq 0 \Rightarrow f_{A|L}(a|l)>0 \,\, \forall (a,l) \in E,
\end{align}
with $E$ the set of pairs $(a,l)$ which are seen in the estimand. For a dynamic scheme, for example, which intervenes on the dosage of a medication with 5 mg for women and 3 mg for men, the positivity assumption needs to be fulfilled for the pairs $(3, \text{man})$ and $(5, \text{woman})$, as that is what is used in the estimation for each group.

\subsubsection{This Equals MTP Positivity}

For a single time point, the definition of the positivity assumption for longitudinal modified treatment schemes is given by 

\begin{center}
``$\text{if } (a_t, h_t) \in \operatorname{supp}\{A_t, H_t\}\text{ then }(\mathbbm{d}(a_t, h_t), h_t) \in \operatorname{supp}\{A_t, H_t\} \text{ for } t \in {1, . . . , \tau }$'' \cite[p. 849]{diaz2023nonparametric} % only in my preprint, I can't access the paper?
\end{center}

which in the single time point case reduces to
\begin{align}
(a^\text{obs}, l) \in \mathrm{supp}\{A, L\} \Rightarrow (\mathbbm{d}(a^\text{obs}, l), l) \in \mathrm{supp}\{A, L\},
\end{align}
where we renamed $h$ to $l$ and $a$ to $a^\text{obs}$, the observed value of $A$, in order to comply with the notation used in this paper. Using the definition of support $\operatorname{supp}(f) := \{ x \in X \,:\, f(x) \neq 0\}$, this can be rewritten to
\begin{align}
f_{A,L}(a^{\text{obs}},l) \neq 0 \Rightarrow f_{A,L}(\mathbbm{d}(a^{\text{obs}}, l)|l) \neq 0 \,\,\forall a^{\text{obs}},l.
\end{align}
And since densities cannot be negative, this leads to 
\begin{align}
f_{A,L}(a^{\text{obs}},l) \neq 0 \Rightarrow f_{A,L}(\mathbbm{d}(a^{\text{obs}}, l)|l) > 0 \,\,\forall a^{\text{obs}},l.
\end{align}
This can be rewritten as
\begin{align}
f_{A,L}(a^{\text{obs}},l) \neq 0 \Rightarrow f_{A,L}(a|l) > 0 \,\,\forall (a=\mathbbm{d}(a^{\text{obs}}, l),l) \in E.
\end{align}

This shows the only difference in the formulation of the positivity assumption for a static treatment and an MTP is that $a^{\text{obs}}$ is included on the left hand side, as in MTP schemes the intervention can depend on this observed value. The additional $a^{\text{obs}}$ is a relaxation of the positivity assumption (like any additional $l$ would relax it as well) as this can only reduce the density on the left hand side: $f_{A,L}(a^{\text{obs}},l) = f_L(l) \cdot f_{A|L}(a^{\text{obs}}|l) \leq f_{L}(l).$ Compared to a static scheme, the MTP also has the advantage that not every observation needs to receive the same intervention, which also leaves room for the researcher to circumvent positivity issues.
These two reasons are why switching to an MTP can help if the originally considered intervention scheme cannot be estimated due to positivity problems.

%This is especially true for small shifts, since $a^{\text{obs}}$ will by definition be close to at least one observed data point and assuming the data is somewhat structured into ``areas with data support'' and ``areas without data support'' a small shift will at the most move observations a short distance out of the supported areas.
%One the other hand, a naive shift (adding a fixed value to $a^{\text{obs}}$) will always lead to a need for extrapolation for finite samples. This follows from the fact that the observation with the largest $a^{\text{obs}}$ in the sample is increased and therefore shifted into an area outside of the data range. 

\clearpage

\section{The Curse Of Dimensionality}\label{sim2}

\begin{wrapfigure}{r}{0.55\textwidth}
    \includegraphics[width=0.55\textwidth]{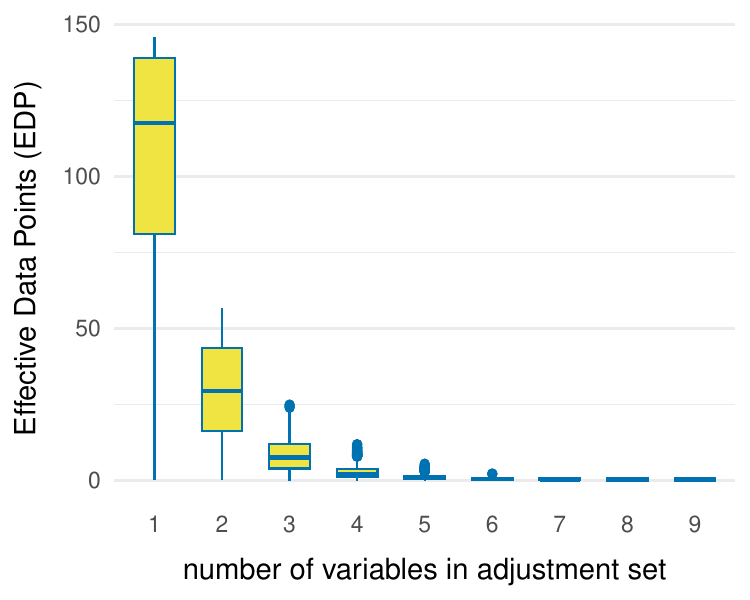}
    \caption{Simulation data showing the change in $EDP$ for a higher dimensionality of the adjustment set.}
    \label{sim2_img}
\end{wrapfigure}
High-dimensional data present a difficulty in estimation which is reflected in the Effective Data Points. In order to illustrate the behavior of the diagnostic for higher dimensional adjustment sets, we present an additional simulation here. The data is simulated as follows:
\begin{align*}
A &\sim \mathrm{Normal}(0,1) \\
L_p &\sim \mathrm{Normal}(0,1),
\end{align*}
with $p=0,...,9$ and $P$, the number of variables included in the adjustment set, incremented for every calculation of the diagnostic. The intervention strategy used for the diagnostic was a simple static treatment of $A=0$ for all 1000 observations.
Figure \ref{sim2_img} shows that the EDP drops drastically with a higher dimensionality of the adjustment set. This represents the Curse Of Dimensionality: the more data dimensions there are the more the information is ``spread out'' meaning with more features data points are bound to get more dissimilar from each other. However, this simulation study represents a sort of worst case scenario, where there is no dependence between the dimensions. In realistic scenarios data are expected to be more clustered, e.\,g.\ in human body measurement data one would expect people with longer arms to have longer legs also. Therefore, the real drop in EDP will likely be less dramatic than in this example, but in general higher dimensions lead to lower EDP.

When working with high-dimensional data, this problem is usually dealt with by making additional assumptions on how complex the functional shape of the predictive model can be, for example by limiting the number of interaction effects. If these assumptions are clearly necessary due to the dimensionality of the data compared to the sample size, they should be reflected in the diagnostic. Often, however, these assumptions are left to the algorithms. 
In order to still obtain meaningful insights from the data, we propose two adjustments to the diagnostic. In both cases, the kernel function is applied separately to each dimension (so the one-dimensional Gaussian if a multivariate Gaussian is utilized), and then the results are combined in a modified way.
The \textit{minimum variant} prevents any single variable from too drastically reducing the overall diagnostic. Specifically, we cap the diagnostic value for each dimension $EDP_p$ at a certain minimum, and then compute the overall EDP as
\[
\text{EDP} = \prod_{p=1}^{P} \max(\text{EDP}_p, \text{minval}_p)
\]
Here, $minval_p$ is a minimum value that can be chosen based on how important variable $p$ is expected to be. For example, for the most important confounders no minimum value is set, while for a confounder which is expected to only have a minor influence on treatment and outcome, a minimum value of $0.5$ could be set. If it's unclear how to set these values, because there is little information on the importance of each feature, the second method may be preferable.
Instead of using a product, the \textit{harmonic mean variant}
combines the per-dimension diagnostics using the harmonic mean:
\[
\text{EDP} = \left( \frac{1}{P} \sum_{p=1}^{P} \frac{1}{\text{EDP}_p} \right)^{-1}
\]
This has the effect of reducing the impact of very low values without requiring subjective minimum thresholds.

%C2 other options for standard variation
%C3 (sim)

%c3 ordinal/cat variables
%c7 kernel, h->0, and h->1
%c9 uniform kernel

\clearpage

\section{Additional True Functional Shapes For Simulation}\label{addsim}

\subsection{Polynomial}

In this section the results for a data-distribution as given in the main text, but without the flattening of the edges, is given.

\subsubsection{Causal Dose Response Curve and  Naive Shift Response Curve}

\begin{figure}[H]
    \centering
    \begin{subfigure}[t]{\textwidth}
        \centering
    \includegraphics[width=\textwidth]{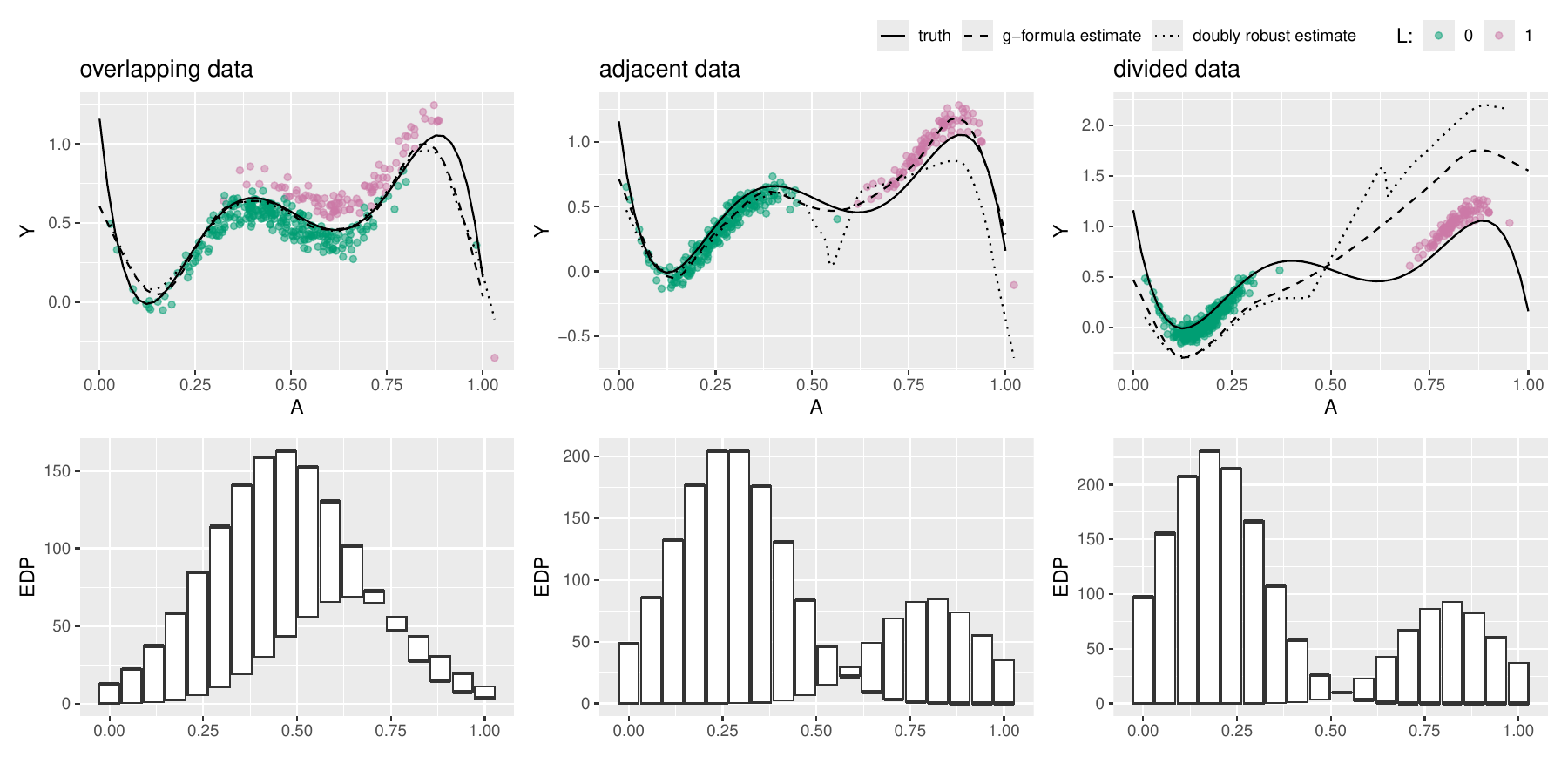}
    \caption{Dose response curves with differing levels of sparsity}
    \label{results:cdrc_2}
    \end{subfigure}
    \begin{subfigure}[t]{\textwidth}
        \centering
    \includegraphics[width=\textwidth]{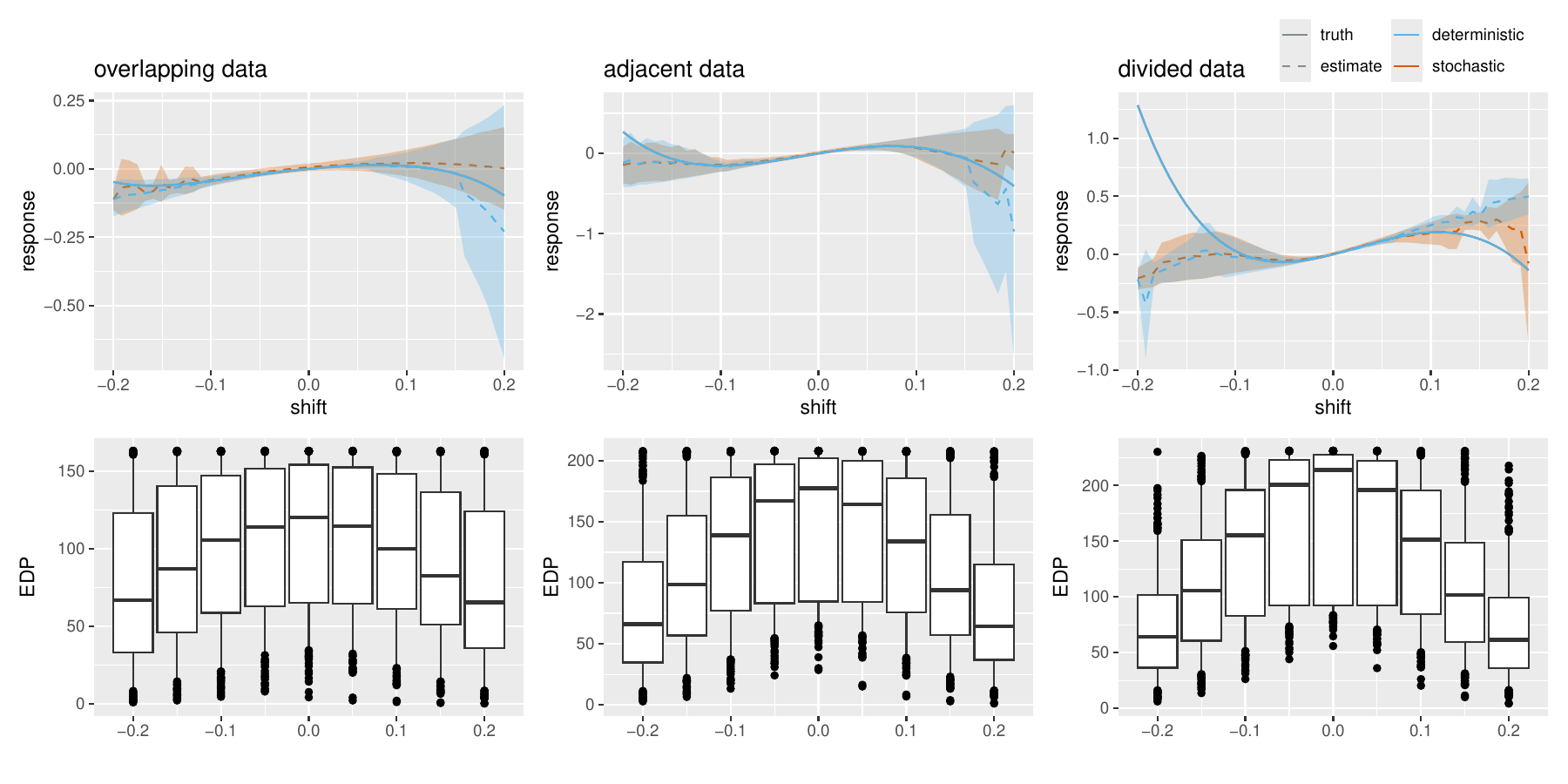}
    \caption{shift response curve}\label{results:src_2}
    \end{subfigure}
    \caption{Simulations for CDRC and naive shift response curve}
\end{figure}

The results for the CDRC and the naive shift response curve for the polynomial data are given in Figures \ref{results:cdrc_2} and \ref{results:src_2}.

\clearpage

% \begin{figure}
%     \centering
%     \includegraphics[width=\textwidth]{figures/polynomial_dia_drc.pdf}
%     \caption{Dose response curves with differing levels of sparsity}
%     \label{results:cdrc_2}
% \end{figure}

% \subsubsection{Shift Response Curve}

% \begin{figure}
% \centering
% \includegraphics[width=\textwidth]{figures/polynomial_dia_src.pdf}
% \caption{shift response curve}\label{results:src_2}
% \end{figure}

\subsubsection{Threshold Response Curve and Conditional Shift Response Curve}

\begin{figure}[H]
    \centering
    \begin{subfigure}[t]{\textwidth}
        \centering
\includegraphics[width=\textwidth]{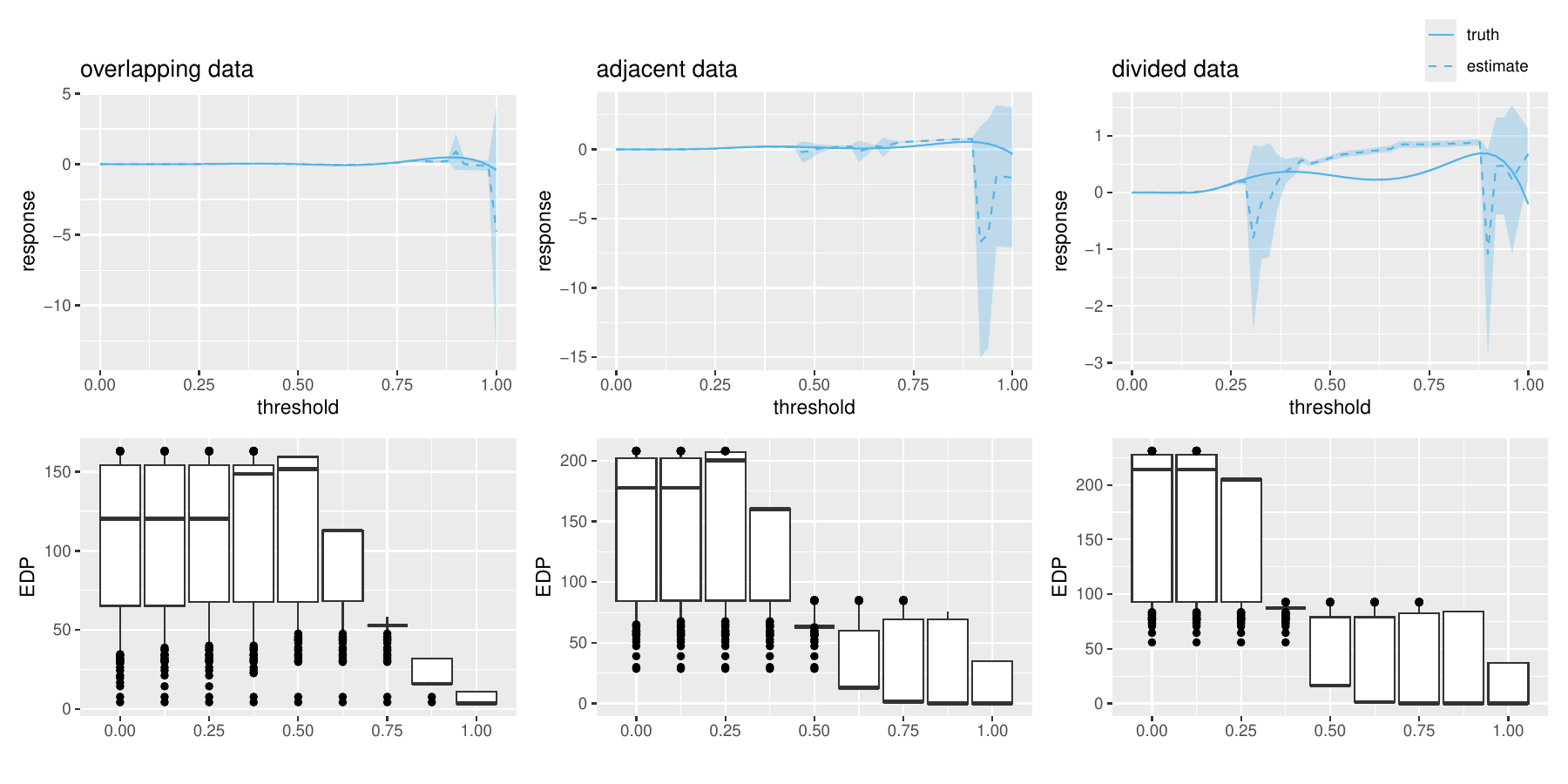}
\caption{threshold response curve}\label{results:trc_2}
    \end{subfigure}
    \begin{subfigure}[t]{\textwidth}
        \centering
\includegraphics[width=\textwidth]{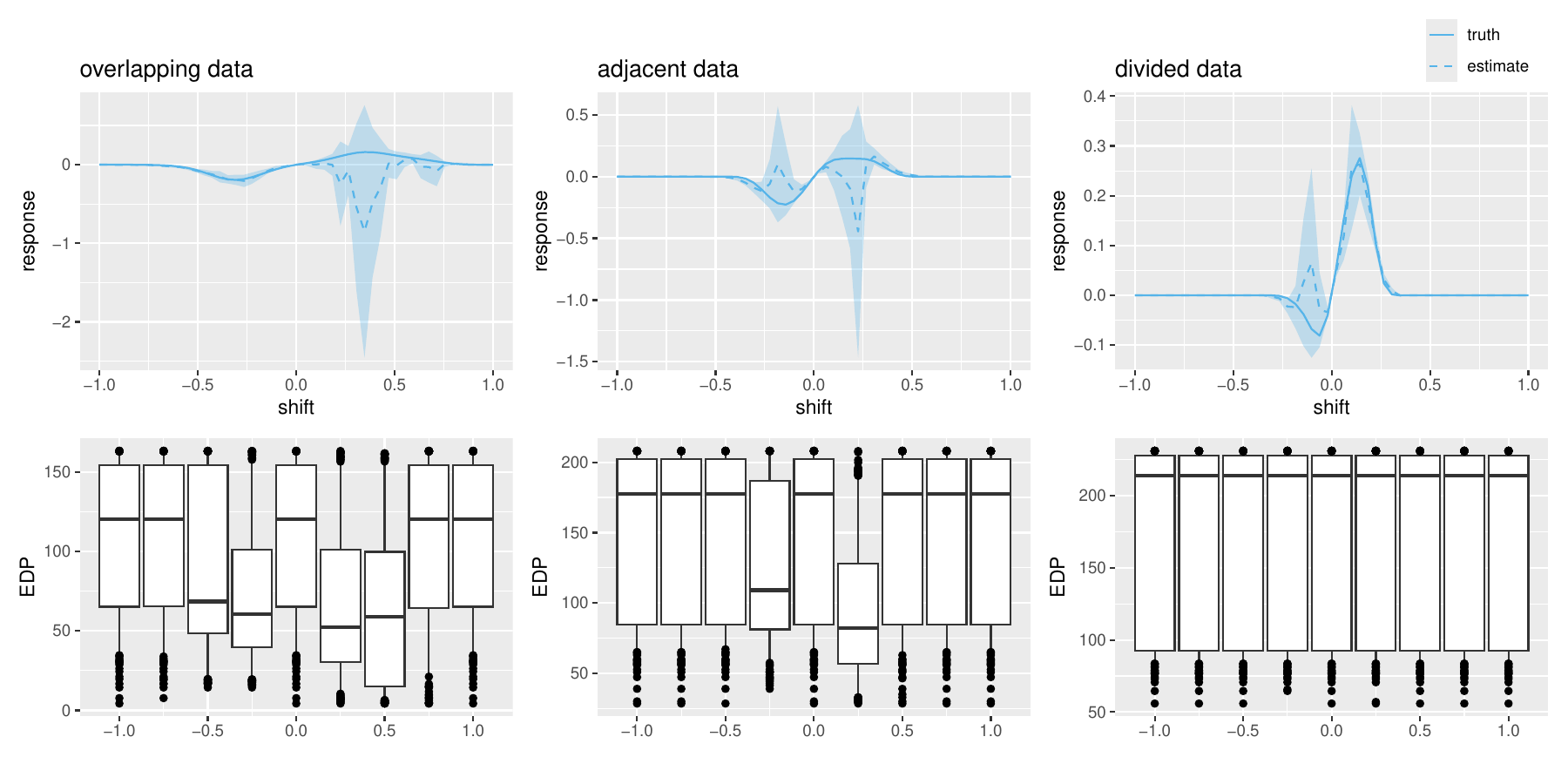}
\caption{conditional shift response curve}\label{results:csrc_2}
    \end{subfigure}

    \caption{Simulations for CDRC and naive shift response curve}
\end{figure}

The results for the threshold response curve and the conditional shift response curve for the polynomial data are given in Figures \ref{results:trc_2} and \ref{results:csrc_2}.

% \begin{figure}
% \centering
% \includegraphics[width=\textwidth]{figures/polynomial_dia_src_d.pdf}
% \caption{dynamic shift response curve (only $L=1$)}\label{results:dsrc_2}
% \end{figure}

% \subsubsection{Threshold Response Curve}

\clearpage

% \begin{figure}
% \centering
% \includegraphics[width=\textwidth]{figures/polynomial_dia_src_th.pdf}
% \caption{threshold response curve}\label{results:trc_2}
% \end{figure}

% \subsubsection{Conditional Shift Response Curve}

% \begin{figure}
% \centering
% \includegraphics[width=\textwidth]{figures/polynomial_dia_src_sh.pdf}
% \caption{conditional \textit{shift} response curve}\label{results:csrc_2}
% \end{figure}

\subsection{Piece-wise Linear}

In this section the results for a data-distribution following a piece-wise linear true dose-response curve, is given.

\subsubsection{Causal Dose Response Curve and  Naive Shift Response Curve}

\begin{figure}[H]
    \centering
    \begin{subfigure}[t]{\textwidth}
        \centering
    \includegraphics[width=\textwidth]{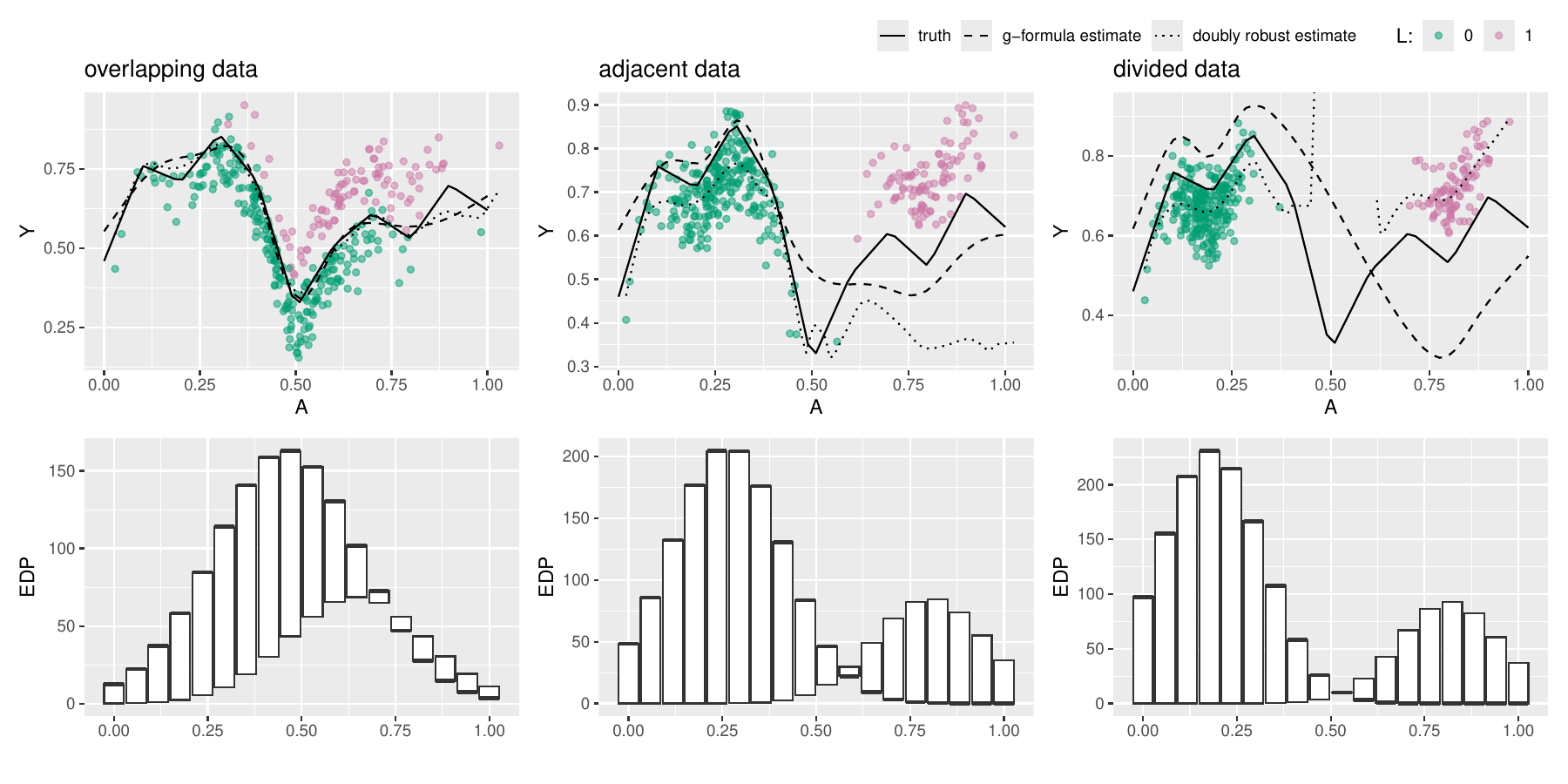}
    \caption{Dose response curves with differing levels of sparsity}
    \label{results:cdrc_3}
    \end{subfigure}
    \begin{subfigure}[t]{\textwidth}
        \centering
    \includegraphics[width=\textwidth]{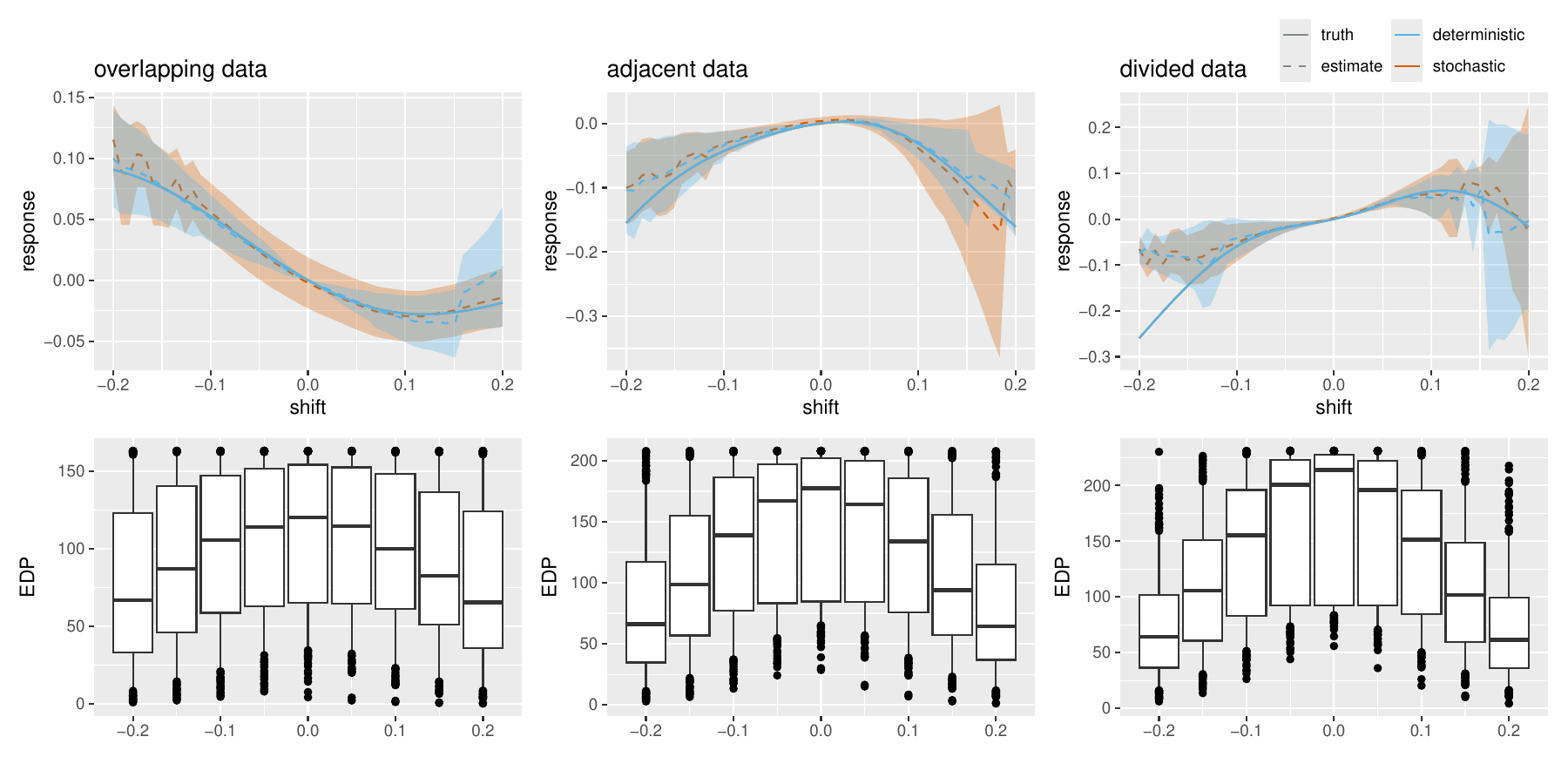}
    \caption{shift response curve}\label{results:src_3}
    \end{subfigure}
    \caption{Simulations for CDRC and naive shift response curve}
\end{figure}

The results for the CDRC and the naive shift response curve for the piece-wise linear data are given in Figures \ref{results:cdrc_3} and \ref{results:src_3}.

\clearpage

\subsubsection{Threshold Response Curve and Conditional Shift Response Curve}

\begin{figure}[H]
    \centering
    \begin{subfigure}[t]{\textwidth}
        \centering
\includegraphics[width=\textwidth]{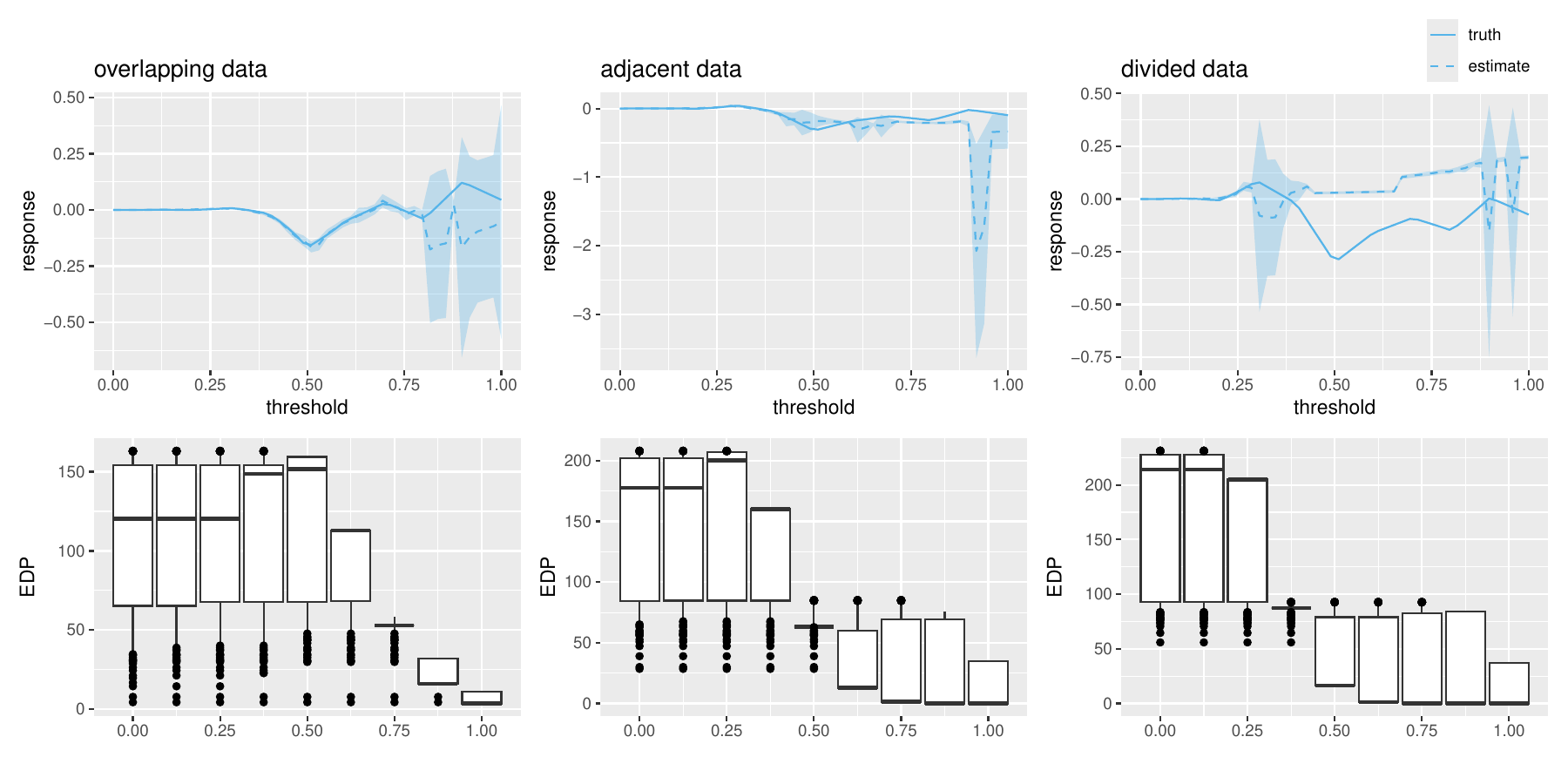}
\caption{threshold response curve}\label{results:trc_3}
    \end{subfigure}
    \begin{subfigure}[t]{\textwidth}
        \centering
\includegraphics[width=\textwidth]{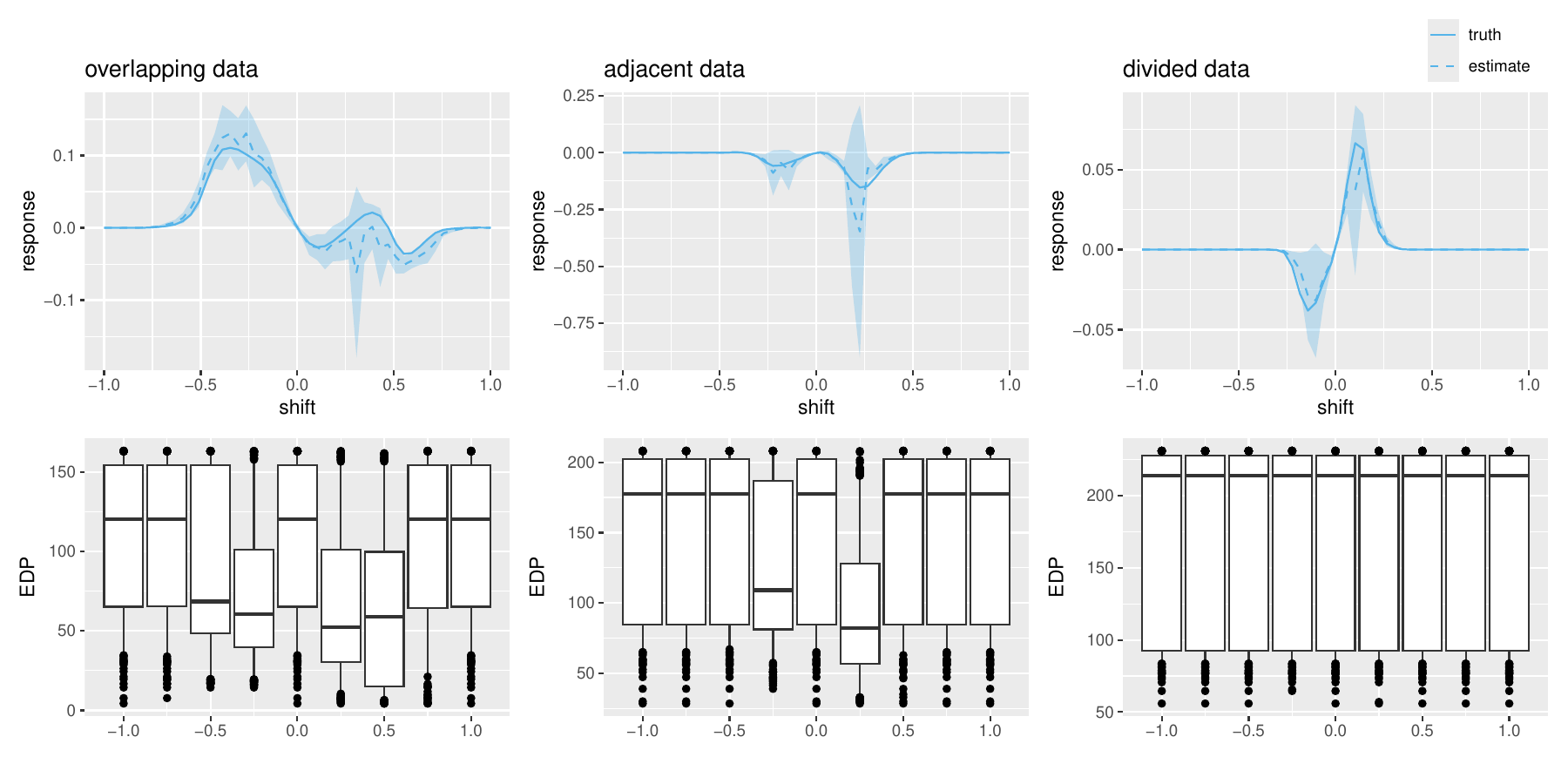}
\caption{conditional shift response curve}\label{results:csrc_3}
    \end{subfigure}

    \caption{Simulations for CDRC and naive shift response curve}
\end{figure}

The results for the threshold response curve and the conditional shift response curve for the polynomial data are given in Figures \ref{results:trc_3} and \ref{results:csrc_3}.

% \subsubsection{Causal Dose Response Curve}

% \begin{figure}
%     \centering
%     \includegraphics[width=\textwidth]{figures/linear-piece2_dia_drc.pdf}
%     \caption{Dose response curves with differing levels of sparsity}
%     \label{results:cdrc_3}
% \end{figure}

% \subsubsection{Shift Response Curve}

% \begin{figure}
% \centering
% \includegraphics[width=\textwidth]{figures/linear-piece2_dia_src.pdf}
% \caption{shift response curve}\label{results:src_3}
% \end{figure}

% \subsubsection{Dynamic Shift Response Curve}

% \begin{figure}
% \centering
% \includegraphics[width=\textwidth]{figures/linear-piece2_dia_src_d.pdf}
% \caption{dynamic shift response curve (only $L=1$)}\label{results:dsrc_3}
% \end{figure}

% \subsubsection{Threshold Response Curve}

% \begin{figure}
% \centering
% \includegraphics[width=\textwidth]{figures/linear-piece2_dia_src_th.pdf}
% \caption{threshold response curve}\label{results:trc_3}
% \end{figure}

% \subsubsection{Conditional Shift Response Curve}

% \begin{figure}
% \centering
% \includegraphics[width=\textwidth]{figures/linear-piece2_dia_src_sh.pdf}
% \caption{conditional \textit{shift} response curve}\label{results:csrc_3}
% \end{figure}

\end{appendices}

\end{document}